\documentclass[reprint,superscriptaddress,aps,prx]{revtex4-2}

\usepackage{amsmath}
\usepackage{amsfonts}
\makeatletter\def\selectlanguage#1{}\makeatother
\usepackage{siunitx}
\usepackage{graphicx}
\usepackage{bm}
\usepackage[caption=false]{subfig}

\usepackage[export]{adjustbox}

\usepackage[version=4]{mhchem}

\newcommand{\ket}[1]{|#1\rangle}

\newcommand{\field}{\text{field}}
\newcommand{\intt}{\text{int}}
\newcommand{\rpa}{\text{RPA}}

\begin{document}

\title{
    Bounds on Bose-Einstein Condensation of Higgs modes in a Superconductor
}

\author{Z. E. Krix} 
\email[]{zeb.krix@unibas.ch}
\affiliation{Department of Physics, University of Basel, Klingelbergstrasse 82, 4056 Basel, Switzerland}
\author{Daniel Loss} 
\affiliation{Physics Department, King Fahd University of Petroleum and Minerals, 31261, Dhahran, Saudi Arabia}
\affiliation{Quantum Center, KFUPM, Dhahran, Saudi Arabia}
\affiliation{RDIA Chair in Quantum Computing}
\affiliation{Department of Physics, University of Basel, Klingelbergstrasse 82, 4056 Basel, Switzerland}

\date{\today}

\begin{abstract}

    Collective modes in a superconductor correspond to fluctuations in the
    amplitude (Higgs mode) or the phase (Goldstone mode) of the order parameter.
    Starting from the BCS Hamiltonian, we derive a microscopic Hamiltonian for
    the coupled Higgs-Goldstone-electron system that contains all interactions
    among these degrees of freedom. The Higgs subsystem is a weakly interacting
    Bose gas: Higgs modes experience a mutual attraction mediated by the virtual
    exchange of Bogoliubov quasi-particle pairs. They also interact with
    external electromagnetic-fields with a coupling constant that we derive
    explicitly. We use this theory to examine whether an optically pumped gas of
    Higgs modes can undergo Bose-Einstein condensation. We identify a density
    window, bounded below by the equilibrium condensation criterion and above by
    collapse due to the attractive interaction, in which condensation can occur
    and we derive the expected density of Higgs-modes due to external pumping.
    This pumping depends crucially on non-parabolic corrections to the electron
    dispersion around the Fermi-energy. We show that the pumped mode density is
    compatible with condensation for realistic material parameters. We suggest
    possible experimental signatures of this non-equilibrium condensate.

\end{abstract}

\maketitle

\section{Introduction}

The Higgs boson of high-energy physics has a condensed-matter analog in
superconductors, known as the Higgs or amplitude mode (throughout this paper
``Higgs mode'' refers to that collective mode of the superconductor, not to its
high-energy analog), first described theoretically in the 1980s
\cite{littlewood_amplitude_1982} and observed directly only within the past
decade \cite{matsunaga_higgs_2013, matsunaga_light-induced_2014}. Experiments
excite the Higgs mode optically in thin (\SI{10}{\nano \meter} to \SI{100}{\nano
\meter}) superconducting films, typically through its non-linear coupling to
light. These observations led to a surge of experimental and theoretical work on
the Higgs mode (see, for example, Ref. \cite{shimano_higgs_2020}), which has
confirmed several of its basic properties: its frequency $2 \Delta$, its
non-linear coupling to light, and its interactions with disorder. This progress
raises a natural question: can a finite-density gas of Higgs modes be maintained
in a superconductor? Doing so requires pumping the modes faster than they decay,
which now appears to be an experimental possibility. If such a gas can be
maintained, what are its properties? For a Bose gas at finite density it is
natural, in particular, to ask whether Bose-Einstein condensation (BEC) can
occur.

The Higgs itself is a collective mode associated with fluctuations in the Cooper
pair density which are supported by the phonon-mediated electron-electron
attraction (the Cooper vertex). The situation is similar to plasmons, which
relate to fluctuations in the charge density and are supported by the Coulomb
electron-electron repulsion. Equivalently, one could think of the Higgs mode as
a propagating fluctuation in the superconducting gap; for useful introductions
to the Higgs mode see Refs. \cite{pekker_amplitudehiggs_2015} and
\cite{shimano_higgs_2020}. Original theoretical work
\cite{littlewood_amplitude_1982} on the Higgs mode provided an explanation of
prior Raman experiments \cite{sooryakumar_raman_1980} on a superconductor with
coexisting charge order. The charge order was crucial in making the Higgs mode
(indirectly) Raman active. Much later, enabled by advances in intense THz
sources, experiments were able to excite the Higgs directly via THz light pulses
in materials with only superconducting order \cite{matsunaga_light-induced_2014}
(THz being the typical frequency range for superconducting gaps). Since then,
the study of the Higgs response in superconductors has progressed steadily.
Different techniques for exciting the Higgs are available: direct excitation
with strong THz light pulses \cite{matsunaga_higgs_2013, huang_discovery_2025},
direct THz excitation in the presence of a supercurrent
\cite{moor_amplitude_2017, yang_lightwave-driven_2019, vaswani_terahertz_2020},
and indirect excitation in Raman-active systems
\cite{glier_non-equilibrium_2025} (that is, in superconductors with coexisting
charge order). These experiments cover a variety of different physical systems,
including $s$-wave superconductors \cite{matsunaga_light-induced_2014},
iron-based superconductors \cite{luo_quantum_2023}, and cuprates
\cite{glier_non-equilibrium_2025}.

Much of the related theory has focused on how light couples to superconductors
and on signatures of collective modes in the resulting optical response (see,
for example, Refs. \cite{schwarz_classification_2020,
mootz_multidimensional_2024}). A conceptual staple of this and related work has
been the pseudo-spin picture of superconductivity developed by Anderson
\cite{anderson_random-phase_1958}. Collective modes are viewed in terms of
precession of pseudo-spins: the $z$-component of pseudo-spin governs the
occupation of electronic states, while the $x$-$y$-components govern the
magnitude and phase of the order parameter (this is described in Ref.
\cite{shimano_higgs_2020} and many other places). Precession (which leads to
collective modes) is induced by a process analogous to that in NMR experiments
\cite{tsuji_theory_2015}. The formal end point of this idea is the application
of a Holstein-Primakoff transform to the pseudo-spin (BCS) Hamiltonian resulting
in explicit creation operators for Higgs and Goldstone modes in terms of
Fermionic operators \cite{tsuchiya_hidden_2018}. This technique is only able to
study collective modes at $q = 0$, however. The present work takes a different
approach. We study the full Hamiltonian of the Higgs mode at all momenta $q \geq
0$: how the modes disperse, how they interact with one another, and how they
couple to their electronic environment. This offers a more general alternative
to the pseudo-spin picture.

Collective modes in other systems are known to condense, but only under external
pumping, since a finite mode density must be maintained. Condensation of
collective modes is therefore a non-equilibrium phenomenon. The longest-studied
example is magnon condensation (and the associated spin superfluidity) in
superfluid \ce{^3He}-B \cite{fomin_long-lived_1984, bunkov_bose-einstein_2008},
where the magnon gas is created in a simple NMR set-up, with the pump operated
either continuously or as a single pulse. The concept of magnon condensation was
introduced to explain extremely long-lived NMR signals and has since led to the
discovery of further effects, including the magnon Josephson effect and new
topological defects (reviewed in Ref. \cite{volovik_twenty_2008}). Once a finite
density is established the conditions for condensation are not stringent because
collective modes generically have small effective masses: critical temperatures
are large and critical densities are small. Magnon condensation has also been
realized in solid-state materials, in particular yttrium iron garnet (YIG)
\cite{demokritov_boseeinstein_2006}. There, the magnon lifetime is much shorter
-- condensates survive for microseconds rather than seconds -- but
thermalization is still possible within this time
\cite{demidov_thermalization_2007}; a central issue has been the sign of the
magnon-magnon interaction, which determines the spatial stability of the
condensate \cite{borisenko_direct_2020}. Beyond collective modes, a large body
of work exists on condensation of quasi-particles such as exciton-polaritons in
semiconductors \cite{byrnes_excitonpolariton_2014}. Those systems lie even
deeper in the non-equilibrium regime, with boson lifetimes sometimes comparable
to the thermalization time.

In the present theoretical work we study a gas of Higgs modes by first deriving
an explicit Higgs Hamiltonian from the BCS reduced Hamiltonian. We do this
following a collective coordinate approach \cite{bardeen_electron-phonon_1955}
and arrive at a Hamiltonian for the combined Higgs-Goldstone-electron system,
including all interactions between these fields. One of our central results is
an explicit expression for the microscopic Higgs-Higgs interaction which is
mediated by exchange of pairs of Bogoliubov quasi-particles and is found to be
attractive. We arrive at a Hamiltonian for the Higgs subsystem, which contains
self-interactions and interactions with Bogoliubov quasi-particles. Some of
these terms are present only at finite Higgs momentum. We also derive terms in
the Hamiltonian which couple the Higgs modes and external electromagnetic
fields; here the square of the vector potential couples to the Higgs field with
a coupling constant that we derive explicitly. The interaction and pumping terms
in the Hamiltonian are most relevant to our study of Bose-Einstein condensation.
In addition, and using a similar collective coordinate method, we obtain an
explicit expression for the Higgs creation operator in terms of electronic
operators at finite momentum, which shows that the Higgs mode involves
pseudo-spin fluctuations not only along $\eta_{x}$ but along all directions
$\eta_{x,y,z}$ (Appendix \ref{sec:EOM}).

We then use this theory to study the problem of Bose-Einstein condensation of
the Higgs mode. Using the known coupling constants from our theory we derive
bounds on the density of Higgs modes within which condensation can occur. As in
prior work on condensation of collective modes, three conditions must be met:
the pump must supply a sufficient density of modes, the gas must remain
spatially stable despite the attractive boson-boson interaction, and the modes
must thermalize within their lifetime. We derive bounds on the first two of
these three points. First, following from our calculation of the Higgs to
external field coupling, we compute the maximum density of Higgs modes, $n_{P}$,
that can be achieved via pumping. Second, following from our calculation of the
attractive Higgs-Higgs interaction vertex, we compute an upper bound on the
Higgs density, $n_{\text{col.}}$, above which a condensate would become
spatially unstable. A lower bound on the Higgs density is supplied by the
equilibrium condition for condition for condensation of a Bose gas
($n_{\text{BEC}}$). We present these bounds in generic form and also map out,
for a series of real materials, the region in the experimental parameter space
where the bounds are consistent with each other and with the possibility for
condensation: $n_{\text{BEC}} < n_{P} < n_{\text{col.}}$.

The paper is organized as follows. Section II develops the collective-coordinate
method used to derive the Hamiltonian of the coupled Higgs-Goldstone-electron
system. Section III presents the resulting Higgs Hamiltonian, including the
Higgs-Higgs self-interaction. Section IV applies this theory to the problem of
BEC and establishes criteria for condensation. Section V suggests two possible
experimental signatures of such a condensate. Technical details, including the
explicit form of the Higgs creation operator, are given in Appendices A--E.

\section{Derivation of the collective-mode Hamiltonian}\label{sec:machinery}

We start from the BCS reduced Hamiltonian with an interaction between
finite-momentum Cooper pairs. The Hamiltonian of the original superconductor is:

\begin{align}
    H = H_{0} + H_{\intt} ,
\end{align}

where

\begin{align}
    \label{eqn:ham0SC}
    H_{0} & = \sum_{k}
    \Psi_{k}^{\dag}
    ( \varepsilon_{k} \eta_{z} + \Delta \eta_{x} )
    \Psi_{k} ,
    \\
    \label{eqn:hamIntSC}
    H_{\intt} & = \frac{U_{0}}{4} \sum_{q}
    \rho_{q}^{+}
    \rho_{-q}^{-} .
\end{align}

We use the following definitions of the Nambu
spinors and density operators:

\begin{align}
    \Psi_{k} & =
    \begin{bmatrix}
        \psi_{k,\uparrow} \\
        \psi_{-k,\downarrow}^{\dag}
    \end{bmatrix} ,
    \\
    \rho_{q}^{i} & =
    \sum_{k}
    \Psi_{k+q}^{\dag}
    \eta_{i}
    \Psi_{k} .
\end{align}

Here $\eta_{0}$ is the identity and $\eta_{x,y,z}$ are the Pauli matrices acting
in Nambu (particle-hole) space; $\eta_{\pm} = \eta_{x} \pm i \eta_{y}$, so that
$\rho_{q}^{\pm}$ in Eq. \eqref{eqn:hamIntSC} gives the Cooper-pair density and
its conjugate. We also assume that the order parameter is oriented along the
$\eta_{x}$ direction. In this picture, the Higgs mode corresponds to
fluctuations in the $\rho_{q}^{x}$ density while the Goldstone mode corresponds
to fluctuations in $\rho_{q}^{y}$ (for small values of the order parameter
phase, $\theta$, the Hamiltonian $H_{0}$ contains $\Delta \theta \eta_{y}$).
While we do \emph{not} do this in the present work, one could also include the
Coulomb interaction and then plasmons would arise as fluctuations in
$\rho_{q}^{z}$. All of these modes are bosonic because the $\rho_{q}^{i}$ are
products of two fermionic fields and are thus effectively bosonic operators.

Dispersion relations for the Higgs and Goldstone modes can be found from the
poles of the self-correlation functions of these density operators. That is,
$\chi_{xx}(\bm{r},t) = \langle \rho^{x}(\bm{r},t) \rho^{x}(0,0) \rangle$
contains the Higgs pole and $\chi_{yy}(\bm{r},t) = \langle \rho^{y}(\bm{r},t)
\rho^{y}(0,0) \rangle$ contains the Goldstone pole. An explicit calculation of
the poles (see, for example, Ref. \cite{littlewood_amplitude_1982}) gives for
the two dispersion relations

\begin{align}
    \omega_{H,q} & = \sqrt{ (2 \Delta)^{2} + c^{2} q^{2} } - i \gamma_{q}
    ,
    \\
    \omega_{G,q} & = cq .
\end{align}

Subscripts $H$ and $G$ will correspond to Higgs and Goldstone modes
respectively for the remainder of this paper. The effective speed of
light and the decay rate are given by

\begin{align}
    c & = v_{F}/\sqrt{3} ,
    \\
    \label{eqn:decayRate}
    \gamma_{q} & = \frac{\pi^{2}}{24} v_{F} q .
\end{align}

The Higgs mode thus has an infinite lifetime at $q = 0$ which rapidly
decreases with $q$. This mode becomes poorly defined for wavelengths
shorter than the coherence length $\xi_{C} = v_{F} / \Delta$, at which
point the typical lifetimes are on the order of $\SI{1}{\pico \second}$
(that is, for $\Delta$ in the meV range).

We derive the Hamiltonian for the collective modes using a simple collective
coordinate approach (as in Ref. \cite{bardeen_electron-phonon_1955}). The need
for such a construction is that in the purely electronic theory the collective
modes exist only implicitly, as poles of the correlation functions $\chi_{xx}$
and $\chi_{yy}$; there is no bosonic operator that could be pumped, counted, or
condensed. The essence of the method is the addition of a pair of collective
fields $\Pi_{H,k}$ and $\Pi_{G,k}$ to the Hamiltonian (through a term
$H_{\text{field}}$) and a pair of constraints which guarantee the equivalence of
the original and transformed Hamiltonians. On the physical subspace selected by
the constraints, $H_{\field}$ contributes nothing and there is no double
counting (the constraints remove exactly as many degrees of freedom as were
added). A canonical transformation (denoted $U_{1}$) is then performed which
brings the collective field part of the Hamiltonian into a harmonically
oscillating form. It does so by entangling each collective coordinate
$\Pi_{i,k}$ with its corresponding fluctuating electronic density operator, so
that the auxiliary oscillator comes to carry the collective component of the
electronic motion. A second canonical transformation (denoted $U_{2}$) then
removes any (3-point) interaction vertices introduced in the previous step (as
in a Schrieffer-Wolff transformation). This step specifically targets
vertices which survive as $q \rightarrow 0$ and would otherwise prevent the
modes from being well defined at long wavelengths, the condition that they cancel
fixes the mode frequencies $\omega_{i,q}$ to the poles of $\chi_{xx}$ and
$\chi_{yy}$. Importantly, the harmonic part of the final Hamiltonian follows
from the random phase approximation (RPA), while interaction terms (for example,
between the Higgs and quasi-particles) follow from non-RPA corrections. We will
denote the two unitary transformations by $U_{1} = e^{i S_{1}}$ and $U_{2} =
e^{i S_{2}}$ and expand in powers of $S_{i}$ up to second order. Our
starting point is then

\begin{align} \label{eqn:hamFull0}
    H & = H_{0} + H_{\intt} + H_{\field} ,
    \\ \label{eqn:hamField1}
    H_{\field} & =
    \sum_{k, i = H, G}
    \frac{1}{2}
    \Pi_{i,k} \Pi_{i,k}^{\dag} ,
    \\ \label{eqn:subsidiaryCond1}
    \Pi_{i,k} \ket{ \psi_{g} } & = 0 .
\end{align}

Here $\ket{\psi_{g}}$ denotes any physical state of the enlarged system, in
particular its ground state; the subsidiary conditions, Eq.
\eqref{eqn:subsidiaryCond1}, select the physical subspace of the enlarged
Hilbert space and thereby ensure the equivalence of $H_{0} + H_{\intt}$ to the
Hamiltonian with $H_{\field}$ added. In perturbation theory, the unitary
transformations are carried out using

\begin{align}
    O' = e^{iS} O e^{-iS} =
    O - i [S, O] - \frac{1}{2} [ S, [S, O] ] +
    \cdots .
\end{align}

We will not require higher orders in this expansion. The form of the first
transformation, $S_{1}$, is fixed by the fact that $\Pi_{H,k}$ should correspond
to fluctuations in $\rho_{q}^{x}$ while $\Pi_{G,k}$ should correspond to
fluctuations in $\rho_{q}^{y}$, as explained below. On the other hand, the form
of $S_{2}$ is fixed by our need to cancel certain interaction vertices.

For the transformation $U_{1} = e^{i S_{1}}$ we introduce the fields, $Q_{i,k}$,
which are canonically conjugate to $\Pi_{i,k}$ as well as the free parameters
$\alpha_{i,k}$ which will be determined below ($\alpha_{i,k}$ must be an even
function of $k$ for $S_{1}$ to be Hermitian), 

\begin{align} \label{eqn:firstTransformation}
    & S_{1} = \sum_{p} \bigl(
    \alpha_{H, p} Q_{H, p} \rho_{p}^{x} +
    \alpha_{G, p} Q_{G, p} \rho_{p}^{y} \bigr) ,
    \\ &
    [ Q_{i, p}, \Pi_{i, k} ] = i \delta_{k,p} .
\end{align}

The auxiliary fields otherwise form mutually commuting canonical pairs:
$[\Pi_{i,p}, \Pi_{j,k}] = [Q_{i,p}, Q_{j,k}] = 0$ for any $i, \ j$ and
$[Q_{i,p}, \Pi_{j,k}] = 0$ for $i \neq j$, and all $\Pi_{i,k}$, $Q_{i,k}$
commute with the electronic operators $\Psi_{k}$. Hermiticity of $H_{\field}$
(Eqn. \eqref{eqn:hamField1}) follows from $\Pi_{i,k}^{\dag} = \Pi_{i,-k}$ and
$Q_{i,k}^{\dag} = Q_{i,-k}$.

Here, we have paired each collective field with its corresponding density
operator. This pairing manifests itself again in the transformed Eq.
\eqref{eqn:subsidiaryCond1},

\begin{align}
\begin{split}
    ( \Pi_{H,k} + \alpha_{H,k} \rho_{k}^{x} ) \ket{ \psi_{g}' } & = 0 ,
    \\
    ( \Pi_{G,k} + \alpha_{G,k} \rho_{k}^{y} ) \ket{ \psi_{g}' } & = 0 ,
\end{split}
\end{align}

where $\ket{\psi_{g}'} = e^{i S_{1}} \ket{\psi_{g}}$ is the transformed physical
state. The form of $S_{1}$ was taken such that the above, transformed condition
would resemble $\Pi_{H,k} + \alpha_{H,k} \rho_{k}^{x} = 0$, etc.

For simplicity, in what follows we present only those terms which are either
quadratic in the collective fields or are interactions containing three field
operators in total (e.g. one collective field and two Fermion fields). This is
the minimal set of terms required to reproduce the Higgs and Goldstone poles.
For the transformation $U_{1}$ we include: $-i[S_{1}, H_{0}]$, $-i[S_{1},
H_{\field}]$, and $-\frac{1}{2}[ S_{1}, [S_{1}, H_{0}] ]_{\rpa}$. Every term but
the final one is a 3-point interaction vertex, the final term is quadratic in
$Q_{i,k}$ and gives the harmonic part of the transformation. For the
transformation $U_{2}$ we also include a term $-i[S_{2},H_{\intt}]_{\rpa}$ which
is zero when computed for $S_{1}$. Note that taking the RPA limit involves
averaging a pair of Fermionic operators over the ground state of the system
(which we take to be the BCS ground state) while enforcing that the momentum
transfer between this pair of operators is zero. Including only the above-listed
terms, the Hamiltonian transformed by $U_{1}$ is given by

\begin{align}\label{eqn:hamAfterS1}
\begin{split}
    H = & H_{0} + H_{\intt} +
    \frac{1}{2} \sum_{q,i=H,G}
    \bigl(
    \Pi_{i,q} \Pi_{i,q}^{\dag} +
    \omega_{i,q}^{2} Q_{i,q} Q_{i,q}^{\dag}
    \bigr)
    \\ & + H_{3} .
\end{split}
\end{align}

The frequencies, $\omega_{i,k}$, in the harmonic part of this Hamiltonian are
free parameters proportional to $\alpha_{i,k}$:

\begin{align}
\omega_{H,q}^{2} & =
\alpha_{H,q}^2 \sum_{k} 2 (\beta^{x}_{k,q})^{2} E_{k,q}^{+},
\\
\omega_{G,q}^{2} & = \alpha_{G,q}^2 \sum_{k} 2 (\beta^{y}_{k,q})^{2} E_{k,q}^{+}.
\end{align}

using parameters $\beta^{i}_{k,q}$ defined in Appendix
\ref{sec:transformations}, Eq. \eqref{eqn:betaFunctions} and $E_{k,q}^{+} =
E_{k} + E_{k+q}$. The series of 3-point interaction terms is given by

\begin{align}\label{eqn:H3Original}
\begin{split}
    H_{3} =
    \sum_{q} \bigl[ &
    \alpha_{H,q} \Pi_{H,q}^{\dag} \rho_{q}^{x} +
    \alpha_{G,q} \Pi_{G,q}^{\dag} \rho_{q}^{y}
    \\ + &
    \alpha_{H,q} Q_{H,q} \delta \rho_{q}^{y} +
    \alpha_{G,q} Q_{G,q} \delta \rho_{q}^{x}
    \\ + &
    \alpha_{H,q} Q_{H,q} \delta \rho_{q}^{0} +
    \alpha_{G,q} Q_{G,q} \delta \rho_{q}^{z} \bigr] .
\end{split}
\end{align}

Note in particular that $S_{1}$ does not introduce any terms like
$Q_{H,k} Q_{G,k}^{\dag}$ which would give mixing between the two
collective modes. In the set of 3-point interaction terms $H_{3}$,
$\delta \rho_{q}^{i}$ is a Fermionic density operator comprising a sum
over $k$ of $\Gamma_{k+q}^{\dag} \eta_{i} \Gamma_{k}$ (for Bogoliubov
spinors, $\Gamma_{k}$ defined in Appendix \ref{sec:transformations}).
The notation $\delta \rho^{i}_{q}$ merely indicates a structural
similarity to $\rho^{i}_{q}$, not a physical relationship.
Specifically, the matrix kernel for each interaction vertex is given
by: $E_{k,q}^{+} \beta_{k,q}^{y} \eta_{x}$ for $\delta \rho_{q}^{x}$,
$- E_{k,q}^{+} \beta_{k,q}^{x} \eta_{x}$ for $\delta \rho_{q}^{y}$, $-
E_{k,q}^{-} \beta_{k,q}^{0}$ for $\delta \rho_{q}^{z}$ and $-i
E_{k,q}^{-} \beta_{k,q}^{z}$ for $\delta \rho_{q}^{0}$. We have used
the notation $E_{k,q}^{\pm} = E_{k} \pm E_{k+q}$, where $E_{k} =
\sqrt{\varepsilon_{k}^{2} + \Delta^{2}}$ is the Bogoliubov
quasi-particle energy, and the functions $\beta_{k,q}^{i}$ defined in
Eq. \eqref{eqn:betaFunctions}.

The strongest interaction terms are those that survive at $q \rightarrow 0$;
this includes all interactions in $H_{3}$ except $\delta \rho_{q}^{z}$ and
$\delta \rho_{q}^{0}$. Collective modes are not well defined in the presence of
these interactions and thus the second unitary transformation, $U_{2} = e^{i
S_{2}}$, exists to cancel them from the Hamiltonian (that is, to cancel
$\rho_{q}^{x,y}$ and $\delta \rho_{q}^{x,y}$ but not $\delta \rho_{q}^{z,0}$).
In applying the transformation $U_{2}$ we consider the same set of terms used in
the $U_{1}$ transformation: $-i[S_{2}, H_{0}]$, $-i[S_{2}, H_{\field}']$, and
$-i[S_{2}, H_{\intt}]_{\rpa}$ (we find that $-\frac{1}{2}[ S_{2}, [S_{2}, H_{0}]
]$ cancels against other non-RPA terms). Note that $H_{\text{field}}'$ is the
entire harmonic part of Eqn. \ref{eqn:hamAfterS1}. We find the following
explicit form for $S_{2}$:

\begin{align}\label{eqn:S2GeneralForm}
    S_{2} =
    \sum_{q, i = H,G} \bigl(
    \Pi_{i, q}^{\dag} A_{i, q} +
    Q_{i, q} B_{i, q} \bigr) ,
\end{align}

where the density operators $A_{i,q}$ and $B_{i,q}$ were found to be

\begin{align}
    A_{H,q} = &
    - \alpha_{H,q}
    \sum_{k}
    \frac
    {\beta^{x}_{k,q} E^{+}_{k,q}}
    {\omega_{H,q}^{2} - (E^{+}_{k,q})^{2}}
    \Gamma_{k+q}^{\dag}
    \eta_{y} \Gamma_{k} ,
    \\
    A_{G,q} = &
    + \alpha_{G,q}
    \sum_{k}
    \frac
    {\beta^{y}_{k,q} E^{+}_{k,q}}
    {\omega_{G,q}^{2} - (E^{+}_{k,q})^{2}}
    \Gamma_{k+q}^{\dag}
    \eta_{x} \Gamma_{k} ,
    \\
    B_{H,q} = &
    + \alpha_{H,q}
    \sum_{k}
    \frac
    {\beta^{x}_{k,q} (E^{+}_{k,q})^{2}}
    {\omega_{H,q}^{2} - (E^{+}_{k,q})^{2}}
    \Gamma_{k+q}^{\dag}
    \eta_{x} \Gamma_{k} ,
    \\
    B_{G,q} = &
    + \alpha_{G,q}
    \sum_{k}
    \frac
    {\beta^{y}_{k,q} (E^{+}_{k,q})^{2}}
    {\omega_{G,q}^{2} - (E^{+}_{k,q})^{2}}
    \Gamma_{k+q}^{\dag}
    \eta_{y} \Gamma_{k} .
\end{align}

Cancellation of the $Q_{i,q}$ interaction vertices in $H_{3}$ (against
$-i[S_{2}, H_{0} + H_{\field}']$) fixes the form of $S_{2}$ (as expressed
above), while cancellation of the $\Pi_{i,q}$ interactions (against $-i[S_{2},
H_{\intt}]_{\rpa}$) leads to the following additional conditions:

\begin{align}
    U_{0} \sum_{k}
    \frac{(\beta_{k,q}^{x})^{2} E_{k,q}^{+}}
    {\omega_{H,q}^{2} - (E_{k,q}^{+})^{2}}
    =
    U_{0} \sum_{k}
    \frac{(\beta_{k,q}^{y})^{2} E_{k,q}^{+}}
    {\omega_{G,q}^{2} - (E_{k,q}^{+})^{2}}
    = 1 .
\end{align}

This equation is our dispersion relation; it determines the mode frequencies
$\omega_{i,q}$ which were free parameters up until this step. Expanding the
$\beta_{k,q}^{i}$ functions (using Eq. \eqref{eqn:betaFunctions}) we find

\begin{align}\label{eqn:poleCondition}
    U_{0} \sum_{k}
    \frac{E + E'}{2 E E'}
    \frac{E E' + \varepsilon \varepsilon' \pm \Delta^2 }{\omega_{i,q}^{2}
    - (E + E')^{2}}
    = 1 ,
\end{align}

where primed quantities are evaluated at $k+q$ and un-primed quantities at $k$.
For $i = H$ we take the minus sign while for $i = G$ we take the plus sign.
These conditions are identical to the pole conditions for the pair-density
correlation functions $\chi_{xx}(\omega,q)$ and $\chi_{yy}(\omega,q)$ dressed by
the series of ladder diagrams in $H_{int}$ (see, for example, Ref.
\cite{littlewood_amplitude_1982}). A number of things are worth noting about our
solution. (i) It was our exclusion of the $\delta \rho_{q}^{z,0}$ vertices from
the cancellation procedure that allowed us to reproduce the standard Higgs and
Goldstone poles. (ii) These surviving interaction terms have been derived using
a different method in Appendix \ref{sec:EOM}. (iii) We do not reproduce the
Higgs pole in its entirety: our calculation misses an infinitesimal imaginary
part $i 0^{+}$ in the denominator, which is responsible for the decay rate Eq.
\eqref{eqn:decayRate} and cannot arise in our (Hermitian) Hamiltonian
transformed by unitary operators. (iv) The transformation $U_{2}$ does not
introduce any new terms which are quadratic in the collective fields; that is,
it does not modify the harmonic part of the Hamiltonian in any way.

\section{The Higgs Hamiltonian}

To obtain the full Hamiltonian relevant to the Higgs mode we must also include
various non-RPA terms that were excluded from the above calculation. This
includes commutators which give interactions that are quadratic in both the
Higgs and the Fermion fields, specifically: $- i [ S_{2}, H_{3}
]_{\text{non-RPA}}$ and $- \frac{1}{2} [ S_{1}, [ S_{1}, H_{0} ]
]_{\text{non-RPA}}$. The RPA limit of the former term is exactly zero while the
RPA limit of the latter term has already been accounted for in the harmonic part
of our Hamiltonian; all other commutators which lead to terms quadratic in both
Higgs and Fermion fields cancel exactly. These are the terms which could have
led to a modification of the harmonic part of the Hamiltonian if they were
non-zero in the RPA limit (changing, for example, the mode frequencies or
introducing coupling between different collective modes).

We arrive at the following complete Hamiltonian including both RPA and non-RPA
terms. Below, we list only those terms which involve the Higgs field and drop
the subscript $H$ from $Q$, $\Pi$, and $\alpha$:

\begin{align}\label{eqn:fullHiggsHam}
\begin{split}
    H_{H} = \
    \frac{1}{2}
    &
    \sum_{q} \bigl(
    \Pi_{q} \Pi_{q}^{\dag} +
    \omega_{q}^{2} Q_{q} Q_{q}^{\dag} \bigr)
    %%%%%%%%%%%%%%%%%%%%%%%%%%%%%%%%%%%%%%
    \\ + &
    \sum_{q,k} Q_{q}
    \Gamma_{k+q}^{\dag} \eta_{0} \Gamma_{k}
    (-i \alpha_{q} E_{k,q}^{-} \beta_{k,q}^{z})
    %%%%%%%%%%%%%%%%%%%%%%%%%%%%%%%%%%%%%%
    \\ + &
    \sum_{q',q,k}
    Q_{q} Q_{q'}
    \Gamma_{k+q'+q}^{\dag}
    \eta_{x}
    \Gamma_{k}
    F_{QQ}^{x}(k,q,q')
    %%%%%%%%%%%%%%%%%%%%%%%%%%%%%%%%%%%%%%
    \\ + &
    \sum_{q',q,k}
    Q_{q} Q_{q'}
    \Gamma_{k+q'+q}^{\dag}
    \eta_{z}
    \Gamma_{k}
    F_{QQ}^{z}(k,q,q')
    %%%%%%%%%%%%%%%%%%%%%%%%%%%%%%%%%%%%%%
    \\ + &
    \sum_{q',q,k}
    \Pi_{q}^{\dag} \Pi_{q'}^{\dag}
    \Gamma_{k+q'+q}^{\dag}
    \eta_{x}
    \Gamma_{k}
    F_{\Pi\Pi}^{x}(k,q,q')
    %%%%%%%%%%%%%%%%%%%%%%%%%%%%%%%%%%%%%%
    \\ + &
    \sum_{q',q,k}
    \Pi_{q}^{\dag} \Pi_{q'}^{\dag}
    \Gamma_{k+q'+q}^{\dag}
    \eta_{z}
    \Gamma_{k}
    F_{\Pi\Pi}^{z}(k,q,q')
    .
\end{split}
\end{align}

The vertex functions are given by:

\begin{align}\label{eqn:4pointInteractionVerticesQQx}
    \begin{split}
        &F_{QQ}^{x}(k,q,q')
        \\ & = \
        \frac{1}{2}\alpha_q \alpha_{q'}
        \left(
            \beta_{k+q',q}^{z}\beta_{k,q'}^{x}E^{+}_{k,q'}
            +
            \beta_{k,q}^{z}\beta_{k+q,q'}^{x}E^{+}_{k+q,q'}
        \right) ,
    \end{split}
\end{align}

\begin{align}\label{eqn:4pointInteractionVerticesQQz}
    \begin{split}
        &F_{QQ}^{z}(k,q,q')
        \\ & = \
        -\frac{1}{2}\alpha_q \alpha_{q'}
        \left(
            \beta_{k+q',q}^{x}\beta_{k,q'}^{x}E^{+}_{k,q'}
            +
            \beta_{k,q}^{x}\beta_{k+q,q'}^{x}E^{+}_{k+q,q'}
        \right) ,
    \end{split}
\end{align}

\begin{align}\label{eqn:4pointInteractionVerticesPiPix}
    \begin{split}
        &F_{\Pi\Pi}^{x}(k,q,q')
        \\ & = \
        \alpha_q \alpha_{q'}
        \left(
            \frac{
                \beta_{k+q',q}^{x}\beta_{k,q'}^{x}E^{+}_{k,q'}
            }{
                \omega_{H,q}^{2} - \left(E^{+}_{k+q',q}\right)^{2}
            }
            +
            \frac{
                \beta_{k,q}^{x}\beta_{k+q,q'}^{x}E^{+}_{k+q,q'}
            }{
                \omega_{H,q}^{2} - \left(E^{+}_{k,q}\right)^{2}
            }
        \right) ,
    \end{split}
\end{align}

\begin{align}\label{eqn:4pointInteractionVerticesPiPiz}
    \begin{split}
        &F_{\Pi\Pi}^{z}(k,q,q')
        \\ & = \
        -\alpha_q \alpha_{q'}
        \left(
            \frac{
                \beta_{k+q',q}^{z}\beta_{k,q'}^{x}E^{+}_{k,q'}
            }{
                \omega_{H,q}^{2} - \left(E^{+}_{k+q',q}\right)^{2}
            }
            +
            \frac{
                \beta_{k,q}^{z}\beta_{k+q,q'}^{x}E^{+}_{k+q,q'}
            }{
                \omega_{H,q}^{2} - \left(E^{+}_{k,q}\right)^{2}
            }
        \right) .
    \end{split}
\end{align}

We note that $H_{H}$ in Eqn. \eqref{eqn:fullHiggsHam} is Hermitian under the
conventions $\Pi_{q}^{\dag} = \Pi_{-q}$, $Q_{q}^{\dag} = Q_{-q}$, with
$\alpha_{q}$ and $\omega_{H,q}$ even in $q$.

In words, the Hamiltonian contains (from top to bottom in Eqn.
\ref{eqn:fullHiggsHam}): (i) a term which governs harmonic motion of the
collective modes, (ii) a 3-point interaction vertex involving a single Higgs
mode and two Fermions, (iii) 4-point interaction vertices involving two Higgs
modes and two Fermions. Importantly, the 3-point interaction disappears in the
limit that the Higgs momentum goes to zero (due to the factor $E_{k,q}^{-}$)
while the 4-point interactions all survive in this limit. Note that it is
possible to derive the same 3-point vertex from direct solution of the equations
of motion; this is presented in Appendix \ref{sec:EOM}.

We can also find terms which are linear in the Higgs field and quartic in the
Fermionic fields. The only non-zero term of this form comes from $-i[S_{2} +
S_{1}, H_{\intt}]_{\text{non-RPA}}$. These are presented in appendix
\ref{sec:summaryOfTerms}.

\subsection{Higgs self-interaction}

The final ingredient in our description of the Higgs gas is the interaction
between the Higgs modes themselves. The main part of this self-interaction comes
from the 4-point Higgs-Fermion interaction terms; the 3-point interactions will
give a contribution scaling like $q^{4}$ and can thus be neglected. It turns out
that this interaction comes entirely from the virtual exchange of a \emph{pair}
of Bogoliubov quasi-particles with nearly opposite momenta and that it is
\emph{attractive}. We compute the Higgs-Higgs self-interaction by accounting for
the 4-point interaction vertices up to second order in perturbation theory. The
full calculation is presented in Appendix \ref{sec:HHInteraction}. In this
calculation, terms which mix $\eta_{x}$ and $\eta_{z}$ vertices are zero, as
well as terms containing two $\eta_{z}$ vertices. Of the remaining terms the
contribution from $Q_{q} Q_{q'}$ is dominant and ultimately leads to the
following self-interaction

\begin{align}
    H_{H-H} & =
    \sum_{ \{q_{i}\} }
    U_{H}
    \xi_{q_{1}}^{\dag}
    \xi_{q_{2}}^{\dag}
    \xi_{q_{3}}
    \xi_{q_{4}}
    \delta_{q_{1} + q_{2}, q_{3} + q_{4}},
    \\
    U_{H} & =
    -
    \alpha_{0}^{4}
    N_{0}
    \ln(\Lambda / \Delta) .
\end{align}

Here $\xi_{q}^{\dag}$ ($\xi_{q}$) creates (annihilates) a Higgs mode of momentum
$q$; these bosonic operators are related to $Q_{H,q}$ and $\Pi_{H,q}$ in
Appendix \ref{sec:transformations}.  Furthermore, $\alpha_{0} = \alpha_{H,q=0} =
(\Delta / \Lambda) \sqrt{1 / N_{0}}$, $N_{0}$ is the total per-spin normal-state
density of states at the Fermi level, and $\Lambda$ is the high-energy cutoff of
the pairing interaction (so that $\Delta = 2 \Lambda e^{-1/\gamma_{0}}$ with
$\gamma_{0} = U_{0} N_{0}$). In $U_{H}$ we have approximated the vertex by its
value at zero Higgs momentum. Importantly, $U_{H}$ is negative so the
self-interaction is attractive. Note that the situation is similar for plasmons:
they experience an attractive interaction mediated by exchange of a virtual
particle-hole pair (this is studied, for example, in Ref.
\cite{ruvalds_bound_1976}).

This self-interaction comes from the $\eta_{x}$ vertex in the 4-point
interactions
(Eqns.~\ref{eqn:fullHiggsHam}--\ref{eqn:4pointInteractionVerticesPiPiz}). Recall
that the harmonic terms in the Hamiltonian come from the RPA applied to the
$\eta_{z}$ vertices of the same 4-point interactions. In the RPA the smallness
of our perturbation parameter $\alpha_{H,q}$ is compensated by the large value
of the integral over a single $\eta_{z}$ vertex, whose size is determined by the
linear scaling of the integrand with energy in the UV limit. In the
self-interaction, by contrast: (i) we require two vertices, and only the
$\eta_{x}$-$\eta_{x}$ combination survives, (ii) the surviving vertex does not
scale linearly with energy in the UV limit and (iii) there is an additional
energy denominator coming from second-order perturbation theory. As a result,
the self-interaction is exceedingly weak.

This completes our description of the Higgs mode: its harmonic motion, its
interactions with Bogoliubov quasi-particles, and its interactions with other
Higgs modes. A gas of Higgs modes is thus a weakly interacting Bose gas with
various mechanisms for scattering from Bogoliubov quasi-particles (given
explicitly in Eq. \eqref{eqn:fullHiggsHam}).

\section{The Higgs gas and criteria for its condensation}

We now wish to study a gas of Higgs modes. By this we mean a finite density
ensemble of Higgs modes, or one with a non-zero chemical potential. A generic
property of collective modes is that their chemical potential is exactly zero in
equilibrium. Such a gas is therefore necessarily a non-equilibrium (or
quasi-non-equilibrium) state. Gases of collective modes and quasi-particles have
been studied in a wide array of physical systems: for example, magnons in
superfluid \ce{^3He}-B \cite{bunkov_bose-einstein_2008}, magnons in
yttrium-iron-garnett \cite{demokritov_boseeinstein_2006}, and
excitons/exciton-polaritons in semiconductors
\cite{byrnes_excitonpolariton_2014}. In each of these cases the Bose gas is
established by pumping the system (the pump being what moves the system out of
equilibrium); in \ce{^3He}, for example, an NMR pulse is used as a magnon pump.

One of the central questions in the experiments mentioned above is whether this
gas undergoes Bose condensation. For the Higgs in particular, this would mean
the formation of a new state in which the Higgs creator $\xi_{q=0}^{\dag}$ has a
non-zero expectation value. We derive an explicit form for this operator in
terms fermionic operators in appendix \ref{sec:EOM}. It turns out that for
collective modes the ordinary condition for Bose condensation (for quadratically
dispersing, non-interacting modes), $n \gtrsim 0.17 (m^{*} T)^{3/2}$, is not
very stringent, owing to the generically small effective masses of collective
modes. In our case the Higgs effective mass, $m_{H} = 2 \Delta / c^{2} = 6
\Delta / v_{F}^{2}$, obtained from the small-$q$ expansion of the dispersion
$\omega_{H,q}$ is typically around $m_{H} \sim 10^{-3} m_{e}$, given
superconducting gaps on the order of $\SI{1}{\milli \electronvolt}$ and Fermi
velocities on the order of $\SI{1e6}{\meter \per \second}$. At zero temperature
the critical density vanishes, but for typical temperatures around $T \sim
\Delta$ we have $n_{\text{BEC}} \sim \SI{1e13}{\per \centi \meter \cubed}$.

The simple criterion for condensation presented above assumes a gas with density
$n$ in \emph{thermal equilibrium}. The challenge in obtaining a Bose condensate
is in reaching such an equilibrium (or quasi-equilibrium) state to begin with.
This involves two main stages: (i) a pumping stage whereby the density $n_{P}$
is created, and (ii) a thermalization stage whereby the pumped modes, which are
distributed in some way across a range of momentum states, relax into a Bose
distribution. Both of these processes are described for exciton-polaritons in
Ref. \cite{byrnes_excitonpolariton_2014}.

Various issues arise in connection with points (i) and (ii) given above. First,
both pumping and thermalization require a specific mechanism: one must specify
how the pump creates Higgs modes and which process allows them to scatter into
lower momentum states. Second, Higgs modes have a finite lifetime $\tau_{L}$ and
we should verify whether the processes (i) and (ii) can occur within this
lifetime. A third important point is that condensation involves populating
(macroscopically, over the thermalization time) the $q = 0$ mode and according
to Eq. \eqref{eqn:decayRate} the lifetime of Higgs modes goes like $1/q$. Thus,
we expect any possible condensate to become more stable as it forms.

Another potential issue for an attractive Bose gas is stability; this has been
investigated, for example, in a BEC of $\ce{^7 Li}$ (Ref.
\cite{sackett_boseeinstein_1997}). Attractive Bose condensates are unstable
toward spatial collapse if their total number is higher than a critical value,
which is inversely proportional to the magnitude of the scattering length. In
our case the mutual attraction is weak; however, it still places an upper limit
on the density of Higgs modes that can be reached experimentally. We present the
numerical estimates for the stability criterion below. We address each of these
points - pumping, spatial stability, and thermalization - individually in the
following sections.

\subsection{Pumping}

The simplest possible pumping mechanism creates Higgs modes via a process which
is quadratic in the external light-field. In this section we derive the pumping
term in the Hamiltonian explicitly using the above theory and use this to
estimate the total density of Higgs modes that can be pumped into a
superconducting sample. We start from an external light-field described by the
vector potential $\bm{A}(\bm{r})$, which enters the Hamiltonian in the kinetic
energy ($H_{0}$) via the minimal coupling. Expanding the kinetic energy in
powers of $\bm{A}$ gives a second-order component $[\varepsilon(k - e
\bm{A})]^{(2)} = \{ A_{i}(\bm{r}) A_{j}(\bm{r}), \frac{1}{2} \partial_{k_{i},
k_{j}} \varepsilon(k) \}/2$; we use the anti-commutator here to ensure that the
resulting operator (which contains both $k$ and $\bm{r}$) is Hermitian.
Following Ref. \cite{tsuji_theory_2015} we point the vector potential along an
arbitrary direction $\bm{A} = A \hat{x}$ and assume that $\varepsilon$ is an
isotropic function of the momentum, meaning that our final result is the same
after substituting $\partial_{x}^{2} \varepsilon \rightarrow \nabla^{2}
\varepsilon / 3$. The second derivative is then expanded in powers of
$\varepsilon$: $\nabla^{2} \varepsilon / 3 = c_{0} + c_{1} \varepsilon +
\cdots$. After expanding the kinetic energy to second-order in $\bm{A}(\bm{r})$
and transforming to momentum space we are left with the following external field
term in the Hamiltonian

\begin{align}
    H_{A} =
    \sum_{Q,Q',k}
    A_{Q} A_{Q'}
    \left[ 
        c_{0} +
        \frac{c_{1}}{2} (\varepsilon_{k} + \varepsilon_{k+\widetilde{Q}})
    \right]
    \Psi_{k+\widetilde{Q}}^{\dag}
    \eta_{z}
    \Psi_{k} .
\end{align}

Here $\widetilde{Q} \equiv Q + Q'$ and $A_Q$, the Fourier transformation of
$\bm{A}(\bm{r})$, contains a component which oscillates in time. To introduce
the Higgs mode we follow the procedure already used above; that is, we apply the
unitary transformations $U_{1} = e^{i S_{1}}$ and then $U_{2} = e^{i S_{2}}$ to
the external field Hamiltonian $H_{A}$. For simplicity, we work to first-order
in $S_{1/2}$ and, again, we average over the Fermionic operators in the final
step. The result of this procedure is a term in the Hamiltonian, $H_{P}$, which
accounts for the coupling between the Higgs mode and the external field
$\bm{A}(\bm{r})$:

\begin{align}
    \label{eqn:HamPump}
    H_{P} & = \sum_{Q,Q'}
    A_{Q} A_{Q'}
    I_{\widetilde{Q}}
    \alpha_{\widetilde{Q}}
    \Pi^{\dag}_{-\widetilde{Q}} ,
    \\
    \label{eqn:HamPumpIntegral}
    I_{Q} & =
    \sum_{k}
    \frac{\Delta}{2 E E'}
    \frac{(\varepsilon + \varepsilon')(E + E')}
         {\omega_{Q}^{2} - (E + E')^{2}}
    \left[
        c_{0} + \frac{c_{1}}{2}(\varepsilon + \varepsilon')
    \right] .
\end{align}

In the integral $I_{Q}$ we have used unprimed variables for functions
evaluated at $k$ and primed variables for functions evaluated at $k+Q$.
The amplitude of the pumping term has two parts:
$\alpha_{\widetilde{Q}}$, which is the perturbation parameter contained
in $S_{1/2}$ and $I_{\widetilde{Q}}$ which is the integral arising from
the averaging over Fermionic operators.

Now, we can separate out the time-dependence of the vector potential
$A_{Q} \rightarrow e^{i \Omega t} A_{Q}$ and apply standard
time-dependent perturbation theory in $H_{P}$ to find the transition
rate, $T_{\widetilde{Q}}$, from an initial, vacuum state to the state
with an additional Higgs mode at momentum $\widetilde{Q}$:

\begin{align}
\begin{split}
    T_{\widetilde{Q}} =
    2 \pi &
    \left|
        \sum_{Q,Q'}
        A_{Q}
        A_{Q'}
        \alpha_{\widetilde{Q}}
        I_{\widetilde{Q}}
        \sqrt{ \frac{\omega_{\widetilde{Q}}}{2} }
    \right|^{2}
    \\ & \times
    \delta( \omega_{\widetilde{Q}} - 2 \Omega )
    \rho_{H}(\omega_{\widetilde{Q}})
    d( \omega_{\widetilde{Q}} ) .
\end{split}
\end{align}

Here $\rho_{H}(\omega)$ denotes the density of states of the Higgs modes. The
total transition rate $T$ is obtained by integrating over
$d(\omega_{\widetilde{Q}}) d \Omega_{Q} / 4 \pi$. Using $Q_{0}$ to denote the
solution to $\omega_{Q_{0}} = 2 \Omega$ we find

\begin{align}
    T & =
    2 \pi \Omega \rho_{H}(2 \Omega)
    \alpha_{Q_{0}}^{2}
    I_{Q_{0}}^{2}
    \mathcal{N}_{S} ,
    \\
    \mathcal{N}_{S} & =
    \int \frac{d \Omega_{Q_{0}}}{4 \pi}
    \left|
    \sum_{Q}
    A_{Q}
    A_{Q-Q_{0}}
    \right|^2 .
\end{align}

Multiplying by the lifetime $\tau_{L}$ (the inverse of the decay rate in Eqn.
\ref{eqn:decayRate}) and dividing by the total volume $V$ we arrive at an
estimate for the total density of modes which can accumulate during pumping. For
$\tau_{L}$ we use the lifetime of the mode with momentum $Q = Q_{0}$. We imagine
targeting a particular $|Q_{0}|$ such that the lifetime $\tau_{L}$ is long
enough to perform a desired experiment. Denoting the density due to pumping by
$n_{P}$, we find

\begin{align}
    n_{P} =
    \frac{3^{2} 2^{8}}{\pi^3}
    \frac{\Delta^{2}}{v_{F}^{3} N_{0}}
    \left( \frac{\Delta}{2 \Lambda} \right)^{2}
    I_{Q_{0}}^{2}
    \mathcal{N}_{S} .
\end{align}

We have used $\Omega \sim \Delta$ where possible and $\alpha_{Q_{0}}
\sim \alpha_{0}$. Next, we must estimate the integral $I_{Q_{0}}$
defined in Eq. \eqref{eqn:HamPumpIntegral}. For a purely quadratic
dispersion $\varepsilon = v_{F}(k - k_{F}) + (k - k_{F})^{2}/2m$ with
$c_{0} = 1/m$ and $c_{1} = 0$. In this case, after a simple change of
variables, the integrand is exactly odd in $\varepsilon$ and the
resulting integral $I_{Q_{0}}$ has contributions only from the boundary
of the integration region (around $\varepsilon \sim \Lambda$). The
result is that $I_{Q_{0}} \sim Q_{0}^{2} / \Lambda^{2}$, which is
exceedingly small (this is similar to Ref. \cite{tsuji_theory_2015},
where, working at $Q_{0} = 0$, the coupling to the pump was
found to be exactly zero with only $c_{0}$). We thus look to the $c_{1}$
contribution to $I_{Q_{0}}$; this could come, for example, from a cubic
correction to the dispersion $\varepsilon = v_{F}(k - k_{F}) + (k -
k_{F})^{2}/2m + b_{3}(k - k_{F})^{3}$ in which case $c_{1} = 2 b_{3} /
v_{F}$. With this correction we obtain a logarithmic contribution to
$I_{Q_{0}}$ at $Q_{0} = 0$: $I_{0} = c_{1} \Delta N_{0} \ln( \Delta / 2
\Lambda)$. We thus estimate $I_{Q_{0}}$ by $I_{0}$ (finite $Q_{0}$
corrections will be small and will depend on the precise form of the
electronic dispersion). Note that we account for deviations from a
quadratic dispersion only in our estimate of the density of pumped
modes; here, it makes the difference between zero density and a finite
density, but it can also give corrections to the dispersion of Higgs
modes.

Lastly, the integral $\mathcal{N}_{S}$ can be estimated for a specific
functional form of the external field in coordinate space. We use
$A(\bm{r}) = A_{0} \Theta( R^{2} - x^{2} - y^{2} ) e^{ - z /
\lambda_{L}}$: a circular beam of radius $R$ incident on the external
boundary at $z = 0$. This captures the finite extension of the
light-beam in the $x$-$y$ plane (surface) of the sample as well as the
exponential decay of the field into the bulk ($z > 0$); typical scales
are $R \sim \SI{1}{\milli \meter}$ and $\lambda_{L} \sim \SI{100}{\nano
\meter}$. We find

\begin{align}\label{eqn:NS}
    \mathcal{N}_{S} =
    A_{0}^{4} \frac{V_{P}^{2}}{V^{2}} \frac{1}{16}
    ,
\end{align}

where $V_{P} = 2 \pi R^{2} \lambda_{L}$; $\mathcal{N}_{S}$ is
essentially a geometrical factor that depends on the shape of the external field
(beam) profile. Introducing the dimensionless parameter $\bar{c}_{1} =
k_{F}^{2} c_{1}$, using $A_{0} = E_{0} / \Delta$ with electric field $E_{0}$,
and measuring the density in units of the total electronic density $n_{e}$ we
have

\begin{align}\label{eqn:pumpedDensity}
    \frac{n_{P}}{n_{e}} \approx
    \frac{1}{2}
    \bar{c}_{1}^{2}
    \left[
        \frac{\Delta}{2 \Lambda}
        \ln
        \left(
            \frac{\Delta}{2 \Lambda}
        \right)
    \right]^{2}
\frac{V_{P}^{2} E_{0}^4}{ V k_{F} \varepsilon_{F}^{4}} .
\end{align}

Note that the density of Higgs modes will decrease with sample volume
because these modes are being excited only within a fixed region of
space. Note also that there is no explicit dependence on $Q_{0}$; this
is because at smaller $Q_{0}$ the lifetime increases according to
$Q_{0}^{-1}$ but the density of states of the Higgs modes decreases
according to $Q_{0}^{+1}$ and these changes exactly compensate each
other.

As an example, consider Nb, which was studied recently in relation to the Higgs
mode in Ref. \cite{huang_discovery_2025}; its basic parameters are $k_{F} =
\SI{1.18e8}{\per \centi \meter}$, $\Delta/2\Lambda = 0.032$ and $\lambda_{L} =
\SI{44}{\nano \meter}$, given across Refs. \cite{ashcroft_solid_1976,
parks_superconductivity_2018, parks_superconductivity_2018-1}. The numbers
relevant to the pumped density are: $\Delta / 2 \Lambda = 0.032$, $E_{0}^{4} /
k_{F} \varepsilon_{F}^{4} = \SI{10.6}{\per \centi \meter \cubed}$ for a field of
$\SI{1}{\kilo \volt \per \centi \meter}$, and $V_{P}^{2} / V =
\SI{0.95e-11}{\centi \meter \cubed}$ for $R = \SI{1}{\milli \meter}$ and $V
= (\SI{2}{\milli \meter})^{3}$. Since most materials do not have an isotropic
dispersion and the parameter $\bar{c}_{1}$ is difficult to extract from
experiments, we treat the numbers presented here as a rough estimate of orders
of magnitude. Putting this together we find

\begin{align}
    \frac{n_{P}}{n_{e}}
    \approx
    \SI{5.6e-13}{}
    \bar{c}_{1}^2 f^4 ,
\end{align}

where $f$ is the strength of the field in units of $\SI{1}{\kilo \volt
\per \centi \meter}$; we use $f = 100$ as an estimate of the maximum
experimental value through out the rest of the paper (Ref.
\cite{huang_discovery_2025} quotes $f \approx 70$, for example).
Lastly, we estimate the parameter $\bar{c}_{1}$, which is determined by
the electronic band structure, by $\bar{c}_{1} \sim 1$; this parameter
is difficult to estimate with any accuracy from experimental data, but
the value we use is typical of various model band structures such as
that of a cubic lattice. Our estimate for the maximal density of Higgs
modes that can be pumped (at $f = \SI{1e2}{}$) is thus:

\begin{align}
    n_{P} \approx \SI{5.6e-5}{} n_{e}
    \approx \SI{3.1e18}{\per \centi \meter \cubed} .
\end{align}

This is larger than our maximal estimate of the critical density required for
BEC ($n_{\text{BEC}}$) by 5 orders of magnitude. The value is computed at
a maximal pump strength ($f$) so that by tuning $f$ we could get any Higgs
density from zero up to this value. Note that we have chosen to make this
estimate for a bulk sample rather than a thin slab. We do this to maximize the
lifetime of the modes: in thin slabs the lowest possible momentum in the
direction perpendicular to the slab will be large (on the order of $\Delta /
v_{F}$) and thus modes will be too short-lived to be well-defined.
The price is a dramatically reduced density of modes: thinning the sample
down would increase $n_{P}$ at the cost of a shorter lifetime.

\subsection{Spatial stability}

Ignoring terms which contribute to the finite Higgs lifetime and using the
approximate quadratic form for the Higgs dispersion the Gross-Pitaevskii energy
is

\begin{align} \label{eqn:GPEnergy}
    E = \int d\bm{r} \, \biggl[
    \frac{|\nabla \psi|^{2}}{2 m_{H}} +
    2 \Delta(\bm{r}) |\psi|^{2} +
    \frac{2 \pi a}{m_{H}} |\psi|^{4}
    \biggr] ,
\end{align}

where $\psi(\bm{r})$ is the condensate wave function, normalized to the total
number of Higgs modes, $\int d\bm{r} \, |\psi|^{2} = N$, and $a = - (3/2)
\lambda_{F} ( \Delta^5 / \varepsilon_{F} \Lambda^{4} ) \ln( \Lambda / \Delta )$
is the scattering length for the attractive self-interaction between Higgs modes
(obtained using $\alpha_{0} = (\Delta / \Lambda) \sqrt{1 / V N_{0}}$ and $a =
m_H U_H V / 2 \pi$ with $m_H / m_e^{*} = 3 \Delta / \varepsilon_{F}$ and Fermi
wavelength $\lambda_{F}$).

In a finite-sized system the order parameter is characterized in part
by a length scale, $l$, which determines its spatial extent; this could
be the sample size or other confinement scale, for example. Then the
state with finite $\psi$ corresponds to an energy minimum only for a
total number of particles $N$ satisfying $N < N_{\text{col.}} \sim l /
|a|$ (this condition is derived in Refs. \cite{shi_bose-einstein_1997,
sackett_dynamics_2001}, for example). For a number of particles larger
than this limit compressibility becomes negative and the gas is
unstable towards collapse.

Using our expression for $|a|$, the upper limit on the number density
for Higgs modes is then

\begin{align}\label{eqn:collapseDensity}
    n_{\text{col.}} = \frac{1}{V} \frac{l}{|a|} =
    \frac{1}{V}
    \frac{2 l}{3 \lambda_{F}}
    \frac{\varepsilon_{F}}{\Delta}
    \frac{\Lambda^{4}}{\Delta^{4}}
    \frac{1}{\ln(\Lambda / \Delta)} .
\end{align}

For Nb (with $k_{F} = \SI{1.18e8}{\per \centi \meter}$, and
$\Delta/2\Lambda = 0.032$) we find that $1 / |a| = \SI{1.3e15}{\per
\centi \meter}$; this is the only part of $n_{\text{col.}}$ that is
intrinsic to the material. The remaining factor depends on sample
geometry: $l / V = \SI{12.5}{\per \square \centi \meter}$ if we reuse
the numbers of the previous section ($l \sim R \sim \SI{1}{\milli
\meter}$ and $V = (\SI{2}{\milli \meter})^3$). Putting this together we
estimate $n_{\text{col.}} \approx \SI{1.6e16}{\per \centi \meter
\cubed}$.

There is thus a window of densities between $n_{\text{BEC}}$ and
$n_{\text{col.}}$ in which the basic conditions for Bose-Einstein
condensation can be satisfied: one would have to tune the pumping
strength so that $n_{P}$ sits within this window. We have expressed the
window in terms of the pumping field strength $f$ for the
band-structure parameter $\bar{c}_{1} = 1$: $f = 4$ to $f = 27$ is the
range of field strengths in which condensation is expected. The
parameter $\bar{c}_{1}$ is crucial to the pumped density $n_{P}$ alone,
which scales like $(\sqrt{\bar{c}_{1}}f)^{4}$. Given that experiments
have an upper limit on the pumping field strength, we can estimate for
which values of $\bar{c}_{1}$ this maximal $f$ places $n_{P}$ at one of
the boundaries of the condensation window. For example, the maximal
$n_{P}(f_{max})$ coincides with the upper boundary $n_{\text{col.}}$ at
$\bar{c}_{1} \approx \SI{7e-2}{}$; at this point the entire
condensation window is experimentally accessible. Similarly, the
maximal $n_{P}(f_{max})$ coincides with the lower boundary
$n_{\text{BEC}}$ at $\bar{c}_{1} \approx \SI{2e-3}{}$; at this point
the condensation window is just out of reach. Thus, in this
hypothetical experiment the condensation window is experimentally
accessible for any $\bar{c}_{1} > \SI{2e-3}{}$.

\subsection{Generic Limits on Condensation}

It is also useful to express the condensation window generically, in terms of
the parameters which define the material and the geometry of the experiment. The
two conditions are $n_{P} < n_{\text{col.}}$ and $n_{P} > n_{\text{BEC}}$; we
wish to write these as conditions on $R$, $l$, and $E_{0}$. 

Now we can make use of our expressions for $n_{P}$ (Eqn. \ref{eqn:pumpedDensity}
with $n_{e} = k_{F}^{3} / 3 \pi^{2}$), for $n_{\text{col.}}$ (Eqn.
\ref{eqn:collapseDensity}), and for $n_{\text{BEC}}(T = \Delta) = 2.5 \Delta^{3}
/ v_{F}^{3}$. Thus the upper limit on the pumped density, $n_{P} <
n_{\text{col.}}$, reduces to

\begin{align} \label{eqn:limitCol}
    \frac{R^{4} E_{0}^{4}}{l}
    & <
    \frac{\varepsilon_{F}^{4}}{k_{F} \lambda_{L}^{2}}
    \frac{1}{\bar{c}_{1}^{2}}
    \left[ \frac{\varepsilon_{F}}{\Delta} \right]
    \left[ \frac{\Lambda}{\Delta} \right]^{6}
    \frac{1}
         {\ln^{2} \left( \frac{2 \Lambda}{\Delta} \right)
          \ln \left( \frac{\Lambda}{\Delta} \right)} .
\end{align}

We have left out numerical pre-factors on the order of unity and have placed all
quantities intrinsic to the material on the right-hand-side. Similarly, the
lower limit, $n_{P} > n_{\text{BEC}}$, reduces to

\begin{align}
    \label{eqn:limitBEC}
    \frac{R^{4} E_{0}^{4}}{V}
    & >
    \frac{\varepsilon_{F}^{4} k_{F}}{\lambda_{L}^{2}}
    \frac{1}{\bar{c}_{1}^{2}}
    \left[ \frac{\Delta}{\varepsilon_{F}} \right]
    \left[ \frac{\Lambda}{\varepsilon_{F}} \right]^{2}
    \frac{1}{\ln^{2} \left( \frac{2 \Lambda}{\Delta} \right)} .
\end{align}

\begin{figure}[t]
    \includegraphics[width=\columnwidth]{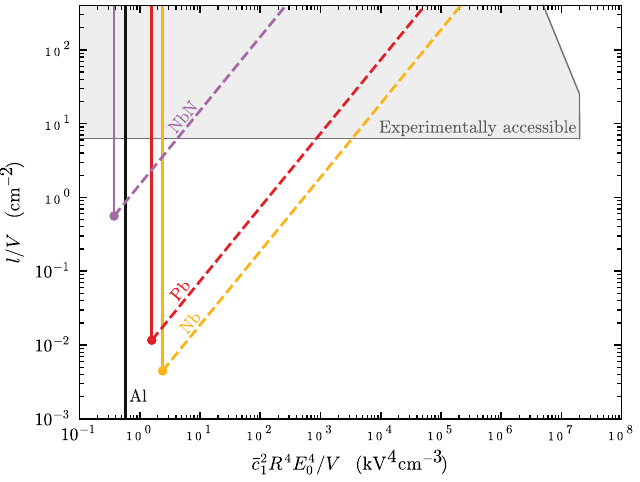}
    \caption{
    Solid lines: Contour $n_{P} = n_{\text{BEC}}$, the lower limit of the
    condensation window. Dashed lines: Contour $n_{P} = n_{\text{col.}}$, the
    upper limit of the condensation window. Colours correspond to different
    materials, each of which are labeled. For each material, condensation is
    possible within the corresponding triangular region. The grey box roughly
    outlines the region the parameter space which is accessible experimentally.
    }
    \label{fig:condensationCriteria}
\end{figure}

These conditions are plotted in Fig. \ref{fig:condensationCriteria} for a number
of different materials (each corresponding to a different colour). Points to the
right of the solid lines satisfy $n_{P} > n_{\text{BEC}}$, while points to the
left of the dashed lines satisfy $n_{P} < n_{\text{col.}}$. The coloured dots
show the points where $n_{\text{BEC}} = n_{\text{col.}}$ and the condensation
window closes. Thus, the condensation window is the set of points within each
triangular shape; points further to the right of this region have a larger
density of Higgs modes. The region for Al contains the plot area to the right of
the black solid line and extends beyond the boundaries of the plot. Figure
\ref{fig:condensationCriteria} also shows the region of parameter space that
experiments have access to (transparent grey area). We defined this using the
rough estimates: $\SI{0.5}{\milli \meter} < R, l < L < \SI{2}{\milli \meter}$
with $V = L^3$ and $E_{0} \leq \SI{100}{\kilo \volt \per \centi \meter}$ and
$\bar{c}_{1} = 1$. Of course, the conditions in Eqns. \ref{eqn:limitBEC} and
\ref{eqn:limitCol} refer only to the possibility of condensation using ratios of
densities; to study the absolute magnitude of the Higgs density it is best to
refer to Eqns. \ref{eqn:pumpedDensity}, \ref{eqn:collapseDensity}, and
$n_{\text{BEC}} \lesssim 0.17 (m_{H} \Delta)^{3/2}$.

\subsection{Thermalization}

Condensation requires that the pumped modes, which are created near the momentum
$Q_{0}$ selected by the pump, redistribute into a Bose distribution within their
lifetime. The dynamics of this process are likely complex and would require more
involved calculations to fully elucidate; we thus deem a complete study of
thermalization outside the scope of this work. Nevertheless, we wish to offer
some comments on this problem. First, multiple scattering channels are available
which could allow the distribution of pumped gas to evolve into a Bose
distribution. (i) The Higgs-Higgs interaction (Eq. \eqref{eqn:fullHiggsHam}) is
available, however, we found this to be weak and anticipate that thermalization
through this channel alone would be slow. (ii) Scattering directly off
Bogoliubov quasi-particles is possible via the 3-point and 4-point vertices of
Eq. \eqref{eqn:fullHiggsHam}. One could tune the availability of Bogoliubov
quasi-particles by varying the temperature. Second, a quantitative estimate of
the thermalization time requires a kinetic treatment of these scattering
processes. Ref. \cite{snoke_population_1989} provides an example of such a
treatment: dynamics of the distribution function are governed by the Boltzmann
equation; all interactions enter via the collision integral. In principle,
the bare ingredients for such a calculation are present in this work already. We
note that Bose-Einstein condensation has been observed in systems where the
lifetime and the thermalization time are roughly equal
\cite{byrnes_excitonpolariton_2014}.

\section{Proposed experiments}

Our suggestions for experiments are based on two fundamental properties
of the Higgs condensate. First, a condensate oscillates with a
spatially uniform frequency set by its chemical potential $\mu$.
Second, the condensate responds to an external ``potential'' given by
the spatial modulation of the equilibrium superconducting gap
$\Delta(\bm{r})$ (see Eqn. \ref{eqn:GPEnergy}). The equilibrium profile
$\Delta(\bm{r})$ can be controlled externally by, for example, a temperature or
magnetic field gradient; in our experimental proposals, we assume that the
sample's environment has been engineered in such a way as to give the desired
profile $\Delta(\bm{r})$. In all cases the profile should vary over length
scales on the order of or larger than $\xi_{C}$.

We will suppose that the probe is an STM tip, which can probe the
density of states of the superconductor via

\begin{align}\label{eqn:STMconductance}
    \frac{dI}{dV}
    &\simeq
    \int dE\,
    \mathrm{DOS}_{\rm tip}(E-eV)\,
    \mathrm{DOS}(E)\,
    \left[-\frac{\partial f(E-eV)}{\partial E}\right].
\end{align}

The simplest way to treat the effect of the condensate on this
measurement is to suppose an oscillating form of the order parameter:

\begin{align}
    \Delta(t)=\Delta_{0}+\delta_{0}\cos(\mu t),
\end{align}

where $\mu$ is independent of position and both $\Delta_{0}$ and $\delta_{0}$
are approximated by their values in the region of space surrounding the tip.
Since the frequency of oscillations, $\mu$, will be in the THz range the $dI/dV$
signal needs to be averaged over time. Thus, we use $\Delta(t) = \Delta_{0} +
\delta_{0}\cos(\mu t)$ in $\mathrm{DOS}(E)$ in Eqn. \ref{eqn:STMconductance} and average
over time. The time average of the density of states is

\begin{align}
    \overline{\mathrm{DOS}}(E)
    &=
    \frac{1}{2\pi}
    \int_{0}^{2\pi} d\phi\,
    \frac{
        E
        \Theta\!\left(E-\Delta_{0}-\delta_{0}\cos\phi\right)
    }{
        \sqrt{E^{2}-\left(\Delta_{0}+\delta_{0}\cos\phi\right)^{2}}
    } .
\end{align}

This integral can be expressed in terms of the complete elliptic
integral of the first kind $\bm{K}(x) = \int_0^{\pi/2} dt / \sqrt{1 - x
\sin^{2}(t)}$. Because of the $\Theta$-function we need to define the
result piece-wise. There are three cases: (i) for $E < \Delta_{0} -
\delta_{0}$ we have $\overline{\mathrm{DOS}}(E) = 0$. (ii) For $\Delta_{0} -
\delta_{0} < E < \Delta_{0} + \delta_{0}$ we have

\begin{align}
    \label{eqn:timeAveragedDOS1}
    \overline{\mathrm{DOS}}(E) =
    \frac{1}{\pi}\sqrt{\frac{E}{\delta_{0}}}\,
    \mathbf{K}\!\left(
    \frac{(E+\delta_{0})^{2}-\Delta_{0}^{2}}
         {4E\delta_{0}}
    \right) .
\end{align}

And (iii) for $E>\Delta_{0}+\delta_{0}$ we have

\begin{align}
    \label{eqn:timeAveragedDOS2}
    \overline{\mathrm{DOS}}(E) =
    \frac{2E}{
    \pi\sqrt{(E+\delta_{0})^{2}-\Delta_{0}^{2}}
    }\,
    \mathbf{K}\!\left(
    \frac{4E\delta_{0}}
         {(E+\delta_{0})^{2}-\Delta_{0}^{2}}
    \right) .
\end{align}

Thus, including a finite $\delta_{0}$ and averaging over time shifts
the singularity in the density of states from $\Delta_{0}$ to
$\Delta_{0} + \delta_{0}$ and converts the singularity from square root
to logarithmic. In addition to this change in the singularity, the
density of states now extends below its usual cut-off $\Delta_{0}$ down
to $\Delta_{0} - \delta_{0}$. Note that the shift in the singularity
and the lower edge of the density of states are both linear in
$\delta_{0}$. The form of the time-averaged density of states near the
singularity is:

\begin{align}
\begin{split}
    \overline{\mathrm{DOS}}(E)
    & \simeq
    \frac{1}{\pi}
    \sqrt{\frac{\Delta_{0} + \delta_{0}}{\delta_{0}}}
    \times
    \\
    &
    \left[
        \log 4
        -\frac{1}{2}
        \log \left(
            \frac{\Delta_{0}}{2\delta_{0}( \Delta_{0} + \delta_{0} )}
            \left| E - \Delta_{0} - \delta_{0} \right|
        \right)
    \right] .
    \label{eqn:logDOS}
\end{split}
\end{align}

Now we consider two different concrete STM experiments that could make use of
this effect. In both, the local amplitude $\delta_{0}(x)$ follows $|\psi(x)|$,
so the pump-induced STM signal maps the mode profile. First, consider a gap
gradient across the sample, $\Delta(x)=\Delta_{0}-Fx$. Since the Higgs potential
is $2\Delta(x)$, the hard wall at the low-gap end forms a triangular well. Its
states are Airy functions with length scale

\begin{align}
    l=(1/4m_{H}F)^{1/3}.
\end{align}

The ground state is localised roughly over a width $2.34l$ on the boundary. For
a sample of length $L= \SI{0.5}{\milli\meter}$ along the direction in which the
$\Delta(x)$ gradient is applied and $F=0.3\Delta_{0}/L$, we find $2.34l\sim
\SI{7.3}{\micro\meter}$ for the confinement length scale (using typical values
$\Delta_{0}=\SI{1}{\milli\electronvolt}$ and $v_{F}=\SI{1e6}{\meter\per\second}$
consistent with Nb, for example). The STM signal should be taken with and
without pumping at either end of the gradient. Then the pump induced shift in
the STM measurement would be visible inside the Airy state and strongly
suppressed outside it. Reversing the gradient should shift the signal to the
opposite side of the sample. Since the STM signal was averaged over time, this
measurement does not directly detect phase coherence, although the precise
profile of $\overline{\mathrm{DOS}}(E)$ that we derived depends a the cosine
form of the gap oscillation underneath the STM tip.

Second, consider a local reduction of the gap over a length $L$ in the middle of
the sample creating a symmetric well of depth $2 \delta \Delta$. If one could
generate a state in this well with at least one node, then comparing STM spectra
at the same positions with and without pumping should show a suppressed
mode-induced change in $dI/dV$ at the node and a finite change on either side.
Since the ground state of the well is nodeless, the first excited state must be
bound and selectively occupied to obtain any node. This state becomes bound when

\begin{align} \label{eqn:nodeCondition}
    L \gtrsim \pi / \sqrt{4 m_{H} \, \delta \Delta} \,.
\end{align}

At $\delta\Delta\sim0.5\Delta_{0}$ (again with typical values $\Delta_{0} =
\SI{1}{\milli\electronvolt}$ and $v_{F} = \SI{1e6}{\meter\per\second}$) this
gives $L \gtrsim \SI{0.6}{\micro\meter}$, so a well of width $L \sim
\SI{1}{\micro\meter}$ could support the state. This length is of order $\xi_{C}$
and is readily resolved by the microscope. If the state is occupied,
$\delta_{0}(x)$ vanishes at its central node. A pump-on--pump-off scan should
then show no condensate-induced line-shape change at the center and a finite
change on both sides. Confinement alone does not populate the excited state, so
this experiment also requires a mode-selective pump or preparation protocol.
Moreover, the time-averaged STM signal measures $|\delta_{0}(x)|$, not its phase
or the common frequency $\mu$. A nodal pattern would support coherent occupation
of one trap mode, but would not by itself prove condensation. A phase-sensitive
or time-resolved measurement would be needed to test the spatially uniform
frequency directly.

\section{Conclusions}

We have presented a theory of the Higgs mode in ordinary
superconductors, including the Higgs-Higgs interaction and the coupling
of the Higgs mode to an external light-field. We showed that the Higgs
modes interact via exchange of virtual pairs of Bogoliubov
quasi-particles, resulting in a weak attraction between modes. Our
theory also includes a series of more generic interactions between
Higgs and Bogoliubov quasi-particles. In these interactions and the
interaction between the Higgs and the light-field we have allowed the
collective modes to take finite momenta. Using this theory, we studied
the possibility of Higgs modes undergoing Bose-Einstein condensation.
Our theory of BEC contains (i) an explicit estimate of the density of
Higgs modes reachable by pumping, and (ii) an explicit estimate of the
upper limit on the Higgs density beyond which collapse occurs due to
the Higgs-Higgs interaction. The lower limit on Higgs density is
defined by the usual BEC criterion in equilibrium and so we have
provided a way to estimate the region of density over which
condensation is possible and a way to estimate where a given experiment
will sit within that window. Our work bridges two currently separate
fields: (i) excitation of Higgs modes and non-linear optical response in
superconductors and (ii) Bose-Einstein condensation of collective modes. Further
work is required on the problem of thermalisation and to pin down expected
experimental signatures of the BEC of Higgs modes. Nevertheless, our results
suggest that Higgs-mode condensation is plausible and motivate its investigation
as a distinct form of Bose–Einstein condensation.

\begin{acknowledgments}
This work was supported by the Georg H. Endress Foundation. DL acknowledges the
Deanship of Research  and the Quantum Center at KFUPM for the support received
under Grant no. CUP25102 and no. INQC2600, respectively.
\end{acknowledgments}

\appendix

\section{Useful transformations and relations}
\label{sec:transformations}

This appendix collects the Bogoliubov transformations, operator identities, and
notational conventions used throughout the main text.

Nambu spinors $\Psi_{k}$ are transformed into the spinors for Bogoliubov
quasi-particle excitations, $\Gamma_{k}$, by the Bogoliubov
transformation:

\begin{align}
    \Gamma_{k} & = M_{k}^{\dag} \Psi_{k} ,
    \\
    M_{k} & =
    \begin{bmatrix}
        u_{k} & -v_{k} \\
        v_{k} &  u_{k}
    \end{bmatrix} .
\end{align}

The Bogoliubov parameters $u_{k}$, $v_{k}$ satisfy

\begin{align}
    \label{eqn:uvRels1}
    u_{k}^{2} & = \frac{1}{2} ( 1 + \varepsilon_{k} / E_{k} ) ,
    \\
    \label{eqn:uvRels2}
    v_{k}^{2} & = \frac{1}{2} ( 1 - \varepsilon_{k} / E_{k} ) ,
    \\
    \label{eqn:uvRels3}
    2 u_{k} v_{k} & = \Delta / E_{k} ,
    \\
    \label{eqn:uvRels4}
    u_{k}^{2} - v_{k}^{2} & = \varepsilon_{k} / E_{k} .
\end{align}

The standard Fermionic commutation relations are

\begin{align}
    \{ \Gamma_{k}^{\alpha}, (\Gamma_{p}^{\beta})^{\dag} \}
    =
    \delta_{\alpha, \beta}
    \delta_{k, p} .
\end{align}

This, of course, also applies to the $\Psi$ operators. A generic
commutator between two density operators is

\begin{align}\label{eqn:baseCommutator}
\begin{split}
    \left[
        \sum_{k}
        \Gamma_{k+q}^{\dag}
        F_{k,q}
        \Gamma_{k}
        ,
        \sum_{p}
        \Gamma_{p+Q}^{\dag}
        G_{p,Q}
        \Gamma_{p}
    \right]
    \\
    =
    \sum_{k}
    \Gamma_{k+q+Q}^{\dag}
    [
        F_{k+Q,q} G_{k,Q} -
        G_{k+q,Q} F_{k,q} 
    ]
    \Gamma_{k} ,
\end{split}
\end{align}

where $F$ and $G$ are $2 \times 2$ matrices. We could have written this
in terms of the $\Psi_{k}$ operators instead.

We will also need to transform density operators from the $\Psi$ basis
to the $\Gamma$ basis (since the interaction Hamiltonian is most
naturally expressed in terms of $\Psi$). This involves a ``mixed''
Bogoliubov transformation since the density operators are defined in
terms of $\Psi_{k+q}^{\dag} \eta_{i} \Psi_{k}$ (with different momenta
$k$ and $k+q$). The transformations are

\begin{align}
    \label{eqn:etaPMTransformed}
\begin{split}
    M_{k+q}^{\dag}
    \eta_{x}
    M_{k}
    & =
    ( 0, \beta_{k,q}^{x}, 0, \beta_{k,q}^{z} ) \cdot \bm{\eta} ,
    \\
    M_{k+q}^{\dag}
    \eta_{y}
    M_{k}
    & =
    ( -i \beta_{k,q}^{0}, 0, \beta_{k,q}^{y}, 0 ) \cdot \bm{\eta} ,
    \\
    M_{k+q}^{\dag}
    \eta_{\pm}
    M_{k}
    & =
    ( \pm \beta_{k,q}^{0}, \beta_{k,q}^{x},
      \pm i \beta_{k,q}^{y}, \beta_{k,q}^{z} ) \cdot \bm{\eta}
    \\
    & =
    \bm{\beta}^{\pm}_{k,q} \cdot \bm{\eta} .
\end{split}
\end{align}

The ``4-vector'' $\bm{\eta}$ just contains the Pauli matrices
$\eta_{i}$ for $i = 0, \ x, \ y, \ z$. The $\beta$ functions are
defined by

\begin{align}
\label{eqn:betaFunctions}
\begin{split}
    \beta_{k,q}^{x} & = u_{k+q} u_{k}- v_{k+q} v_{k}
    \overset{q \to 0}{=}
    \varepsilon_{k} / E_{k} ,
    \\
    \beta_{k,q}^{y} & = u_{k+q} u_{k}+ v_{k+q} v_{k}
    \overset{q \to 0}{=}
    1 ,
    \\
    \beta_{k,q}^{z} & = u_{k+q} v_{k}+ v_{k+q} u_{k}
    \overset{q \to 0}{=}
    \Delta / E_{k} ,
    \\
    \beta_{k,q}^{0} & = u_{k+q} v_{k}- v_{k+q} u_{k}
    \overset{q \to 0}{=}
    0 .
\end{split}
\end{align}

We also give these functions in the limit of zero external momentum
$q$. Note that only $\beta^{0}$ is zero at $q = 0$.

Finally, for the sake of completeness, we note that throughout this
work we use $\xi_{i,q}$ for the annihilation operator of the
Higgs/Goldstone mode (given $i = H$ or $i = G$) and that the canonical
coordinates are defined in terms of these according to

\begin{align}
    Q_{i, p} & =
    \frac{i}{\sqrt{2 \omega_{i,p}}}
    ( \xi_{i, p} - \xi_{i, -p}^{\dag} ) ,
    \\
    \Pi_{i, p} & =
    \frac{\sqrt{\omega_{i,p}}}{\sqrt{2}}
    ( \xi_{i, p}^{\dag} + \xi_{i, -p} ) .
\end{align}

\section{Derivation of Higgs-Higgs interaction}
\label{sec:HHInteraction}

In this appendix we derive the Higgs-Higgs interaction presented in Sec. III. We
denote by $H_{4}$ the 4-point interaction vertices of Eq.
\eqref{eqn:fullHiggsHam}. Accounting for these vertices up to second order
in perturbation theory gives an interaction which scatters the state $\ket{k,p}$
containing two Higgs modes to another state $\ket{k',p'}$ also containing two
Higgs modes. The transition between this initial and final state can occur via
various intermediate states, $\ket{m}$, which always contain the same number of
quasi-particles but are distinguished by their number of Higgs modes (either
zero, 2, or 4). We find that only those terms in the perturbation theory which
contain two $\eta_{x}$ vertices (taken from $H_{4}$, see Eq.
\eqref{eqn:fullHiggsHam}) survive, meaning that the intermediate state always
contains a pair of Bogoliubov quasi-particles. In other words, the Higgs modes
interact with each other via exchange of pairs of Bogoliubov quasi-particles.
The surviving terms are shown diagrammatically below, along with their
corresponding energy denominators, in the limit that all Higgs momenta are
zero:

\begin{align}
    \includegraphics[
        width =0.18\textwidth,
        valign=c,
        raise =0.0\baselineskip
    ]{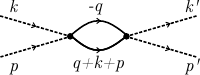}
    \rightarrow
    \frac{1}{2 \omega_{H,0} - 2 E_{q}} ,
\end{align}

\begin{align}
    \includegraphics[
        width =0.15\textwidth,
        valign=c,
        raise =0.0\baselineskip
    ]{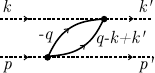}
    \rightarrow
    \frac{-1}{2 E_{q}} ,
\end{align}

\begin{align}
    \includegraphics[
        width =0.15\textwidth,
        valign=c,
        raise =0.0\baselineskip
    ]{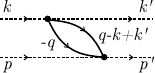}
    \rightarrow
    \frac{-1}{2 E_{q}} ,
\end{align}

\begin{align}
    \includegraphics[
        width =0.15\textwidth,
        valign=c,
        raise =0.0\baselineskip
    ]{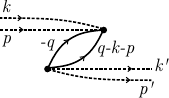}
    \rightarrow
    \frac{-1}{2 \omega_{H,0} + 2 E_{q}} .
\end{align}

Each of these terms needs to be multiplied by the square of the vertex
factor before integrating over $q$. We take the vertex factor from the
quadratic in $Q$ term of Eq. \eqref{eqn:fullHiggsHam} only; this leads
to an integral which diverges logarithmically in the UV cut-off
$\Lambda$, all other terms in Eq. \eqref{eqn:fullHiggsHam} lead to
integrals with no such divergence and are thus neglected. The square of
the vertex factor (again, in the limit of zero Higgs momenta) is

\begin{align}
\begin{split}
    (
        \alpha_{0}^{2}
        \beta_{q,0}^{z}
        \beta_{q,0}^{x}
        E_{q} / \omega_{H,0}
    )^{2}
    =
    \left(
        \alpha_{0}^{2}
        \frac{\Delta \varepsilon_{q}}{\omega_{H,0} E_{q}}
    \right)^{2} .
\end{split}
\end{align}

Thus, at large $\varepsilon_{q}$ the vertex factor goes like
$\varepsilon_{q}^{0}$ and the full integrand with account of the energy
denominators goes like $1 / \varepsilon_{q}$ and is logarithmically
divergent. Completing the integration gives, for the matrix element
which scatters $\ket{k,p}$ to $\ket{k',p'}$, the following

\begin{align}
\begin{split}
    &
    \alpha_{0}^{4}
    \frac{\Delta^{2}}{\omega_{H,0}^{2}}
    N_{0}
    \int d \varepsilon_{q}
    \frac{\varepsilon_{q}^2}{E_{q}^2}
    \left(
        - \frac{1}{E_{q}}
        + \frac{E_{q}}{\omega_{H,0}^{2} - E_{q}^{2}}
    \right)
    \\
    = &
    - 4
    \alpha_{0}^{4}
    \frac{\Delta^{2}}{\omega_{H,0}^{2}}
    N_{0}
    \ln(\Lambda / \Delta) .
\end{split}
\end{align}

\section{Summary of terms in unitary transformations}\label{sec:summaryOfTerms}

Since the calculation presented in section \ref{sec:machinery} contains many
terms, it is useful to enumerate all of the possible terms that we could have
considered. First, the terms that we included to arrive at the Higgs and
Goldstone dispersion relations are given in the table
\ref{tab:restrictedCommutators}. We have used the notation $\Pi^{n} \Psi^{m}$
for a term with $n$ Higgs fields and $m$ fermionic fields. Note that single
commutators come with a prefactor $-i$ and nested commutators come with a
prefactor $-1/2$. If we apply $U_{1}$ up to second order in $S_{1}$ and then
$U_{2}$ up to second order in $S_{2}$ then the set of remaining terms (that is,
all terms not written in table \ref{tab:restrictedCommutators}) are given in
table \ref{tab:allCommutators}.

\begin{table}[h]
\caption{
Terms in the unitary transformations (section \ref{sec:machinery}) that are
required to reproduce the Higgs and Goldstone poles. The first three terms
correspond to the application of $U_{1}$ and the last three terms to the
application of $U_{2}$.
}
\label{tab:restrictedCommutators}
\centering
\begin{tabular}{|| l | l ||}
    \hline
    Term & Form \\ [0.5ex]
    \hline\hline
    $[S_{1}, H_{0}]$                  & $\Pi^{1} \Psi^{2}$ \\
    $[S_{1}, H_{\text{field}}]$       & $\Pi^{1} \Psi^{2}$ \\
    $[S_{1}, [S_{1}, H_{0}] ]_{\rpa}$ & $\Pi^{2} \Psi^{0}$ \\
    \hline
    $[S_{2}, H_{0}]$                  & $\Pi^{1} \Psi^{2}$ \\
    $[S_{2}, H_{\text{field}}']$      & $\Pi^{1} \Psi^{2}$ \\
    $[S_{2}, H_{\intt}]_{\rpa}$       & $\Pi^{1} \Psi^{2}$ \\
    [1ex]
    \hline
\end{tabular}
\end{table}

To arrive at Eqn. \ref{eqn:fullHiggsHam} we accounted for the parts quadratic in
both the Higgs and fermion fields of the terms numbered 3 to 7 in the above
table. Of these, term 3 leads to the quadratic in $Q_{k}$ and $\Gamma_{k}$ part
of Eqn. \ref{eqn:fullHiggsHam}, while term 7 leads to the quadratic in $\Pi_{k}$
and $\Gamma_{k}$ part of the same equation. All other terms which could also
give a form quadratic both in Higgs and fermion fields cancel among themselves
(these are 4 to 6 in the table \ref{tab:allCommutators}).

Note that in table \ref{tab:restrictedCommutators} the terms included for
$U_{1}$ and $U_{2}$ may look asymmetric. This is because $[S_{1},
H_{\intt}]_{\rpa}$ is not included for the first transformation and $[S_{2},
[S_{2}, H_{0} ]]$ is not included for the second transformation. This is not an
issue, however, because the first of these terms is exactly zero, while the
second of these terms cancels (as already described).

Other terms that were entirely left out include terms quartic in the fermion
fields (that is, from terms 5 to 7 above). These give small corrections to the
electron-electron interaction.

Lastly, we present expressions for the terms of the form $\Pi^{1} \Psi^{4}$
involving the Higgs fields: $-i[S_{1}, H_{\intt}]$ and $-i[S_{2}, H_{\intt}]$.
The first of these is

\begin{align}
    -i[S_{1}, H_{\intt}] =
    -i \frac{U_{0}}{2}
    \sum_{q,p}
    \alpha_{H,p} Q_{H,p}
    [ 
        \rho_{q}^{+}   \rho_{p-q}^{z} -
        \rho_{p-q}^{z} \rho_{q}^{-} 
    ].
\end{align}

The second term has a more complicate structure, it is given by

\begin{align}
\begin{split}
    -i[S_{2}, H_{\intt}] =
    \\
    -i \frac{U_{0}}{4} \sum_{p,q}
    & \Pi_{H,p}^{\dag}
        ( 
             \rho_{q}^{+} [A_{H,p}, \rho_{-q}^{-}] +
             [A_{H,p}, \rho_{-q}^{+}] \rho_{-q}^{-}
        ) 
    \\
    & + Q_{H,p}
        ( 
             \rho_{q}^{+} [B_{H,p}, \rho_{-q}^{-}] +
             [B_{H,p}, \rho_{-q}^{+}] \rho_{-q}^{-}
        ).
\end{split}
\end{align}

The internal commutators are given by:

\begin{align}
\begin{split}
    [A_{H,p}, & \rho_{\pm q}^{\pm}] =
    \\ &
    \sum_{k} \Gamma_{k+p \pm q}^{\dag}
    [
        a_{k\pm q, p} \eta_{y} \bm{\beta}_{k,\pm q}^{\pm} \cdot \bm{\eta} -
        a_{k, p} \bm{\beta}_{k,\pm q}^{\pm} \cdot \bm{\eta} \eta_{y} 
    ] \Gamma_{k},
\end{split}
\end{align}

\begin{align}
\begin{split}
    [B_{H,p}, \rho_{\pm q}^{\pm}] =
    - \sum_{k} \Gamma_{k+p \pm q}^{\dag}
    [ &
        E^{+}_{k\pm q, p} a_{k\pm q, p}
        \eta_{x} \bm{\beta}_{k,\pm q}^{\pm} \cdot \bm{\eta} 
    \\ &
        -E^{+}_{k, p} a_{k, p}
        \bm{\beta}_{k + p,\pm q}^{\pm} \cdot \bm{\eta} \eta_{x}
    ] \Gamma_{k}.
\end{split}
\end{align}

Where we have used the notation

\begin{align}
    a_{k,q} = &
    - \alpha_{H,q}
    \frac
    {\beta^{x}_{k,q} E^{+}_{k,q}}
    {\omega_{H,q}^{2} - (E^{+}_{k,q})^{2}}.
\end{align}

\begin{table}[h]
\caption{
Terms in the unitary transformations (section \ref{sec:machinery}) which were
not included in Table \ref{tab:restrictedCommutators}.
}
\label{tab:allCommutators}
\centering
\begin{tabular}{|| l | l | l ||}
    \hline
    \# & Term & Form \\ [0.5ex]
    \hline\hline
    1  & $[S_{1}, H_{\intt} ]_{\text{non-RPA}}$       & $\Pi^{1} \Psi^{4}$ \\
    2  & $[S_{2}, H_{\intt} ]_{\text{non-RPA}}$       & $\Pi^{1} \Psi^{4}$ \\
    3  & $[ S_{1}, [S_{1}, H_{0}] ]_{\text{non-RPA}}$ & $\Pi^{2} \Psi^{2}$ \\
    4  & $[ S_{2}, [S_{2}, H_{0}] ]$                  & $\Pi^{2} \Psi^{2}$ \\
    5  & $[ S_{1}, [S_{1}, H_{\text{field}}] ]$       & $\Pi^{2} \Psi^{2}, \Psi^{4}$ \\
    6  & $[ S_{2}, [S_{2}, H_{\text{field}}'] ]$      & $\Pi^{2} \Psi^{2}, \Psi^{4}$ \\
    7  & $[ S_{2}, H_{3} ]$                           & $\Pi^{2} \Psi^{2}, \Psi^{4}$ \\
    8  & $[ S_{1}, [S_{1}, H_{\intt}] ]$              & $\Pi^{2} \Psi^{4}, \Psi^{6}$ \\
    9  & $[ S_{2}, [S_{2}, H_{\intt}] ]$              & $\Pi^{2} \Psi^{4}, \Psi^{6}$ \\
    10 & $[ S_{2}, [S_{1}, H_{\intt}] ]$              & $\Pi^{2} \Psi^{4}, \Psi^{6}$ \\
    [1ex]
    \hline
\end{tabular}
\end{table}

\section{Equations of motion}\label{sec:EOM}

In this appendix we solve the equations of motion directly, which gives not only
the collective mode dispersions but also an explicit creation operator for the
Higgs and Goldstone excitations expressed in terms of Fermionic operators. The
starting point for the calculation is the same Hamiltonian with a non-zero
pair-field and a contact attraction between electrons (see Eqs.
\eqref{eqn:ham0SC}, \eqref{eqn:hamIntSC}). The equations of motion can be solved
explicitly within the RPA (see Ref. \cite{bardeen_electron-phonon_1955}) and can
also be used to derive the 3-point interaction vertices presented in the main
text (Eq. \eqref{eqn:fullHiggsHam}). This section derives both the dispersion
relations and the 3-point interaction vertices from the equations of motion.

First, work with the Bogoliubov-transformed Nambu spinor $\Gamma_{k} =
M_{k}^{\dag} \Psi_{k}$ rather than with $\Psi_{k}$. Suppose that the
annihilation operator for a generic collective mode is given by

\begin{align}\label{eqn:xiqDef}
    \xi_{q} = \sum_{k}
    \Gamma_{k+q}^{\dag}
    ( \bm{\alpha}_{k,q} \cdot \bm{\eta} )
    \Gamma_{k} ,
\end{align}

where $\bm{\alpha}_{k,q}$ is a vector which is to be determined. The
equation we wish to solve is

\begin{align}\label{eqn:EOM}
    [ \xi_{q}, H_{0} + H_{\intt} ] - \omega \xi_{q} = 0 + \cdots .
\end{align}

The $\cdots$ refers to terms which can be neglected within the RPA.
After making the transformation to the $\Gamma$ operators we have

\begin{align}
    H_{0} = \sum_{k} \Gamma_{k}^{\dag} E_{k} \eta_{z} \Gamma_{k} .
\end{align}

And the interaction Hamiltonian is

\begin{align}
    H_{\intt} =  \frac{U_{0}}{4} \sum_{q} \rho_{q}^{+} \rho_{-q}^{-} ,
\end{align}

with density operators

\begin{align}\label{eqn:sigmaQ}
    \rho_{q}^{\pm} =
    \sum_{k}
    \Gamma_{k}^{\dag}
    M_{k+q}^{\dag}
    \eta_{\pm}
    M_{k}
    \Gamma_{k} ,
\end{align}

given by Eq. \eqref{eqn:etaPMTransformed}. We compute the LHS of Eq.
\eqref{eqn:EOM} by first computing the commutator with $H_{0}$ (making
use of the generic commutation relation Eq. \eqref{eqn:baseCommutator}):

\begin{align}
\begin{split}
    [ \xi_{q}, H_{0} ] & =
    \sum_{k}
    \Gamma_{k+q}^{\dag}
    [
        E_{k} (\bm{\alpha}_{k,q} \cdot \bm{\eta}) \eta_{z} -
        E_{k+q} \eta_{z} (\bm{\alpha}_{k,q} \cdot \bm{\eta})
    ]
    \Gamma_{k}
    \\
    & =
    \sum_{k}
    \Gamma_{k+q}^{\dag}
    [
        i E^{+} ( \alpha^{y}_{k,q} \eta_{x} -
        \alpha^{x}_{k,q} \eta_{y} ) +
          E^{-} \alpha^{z}_{k,q} \eta_{0}
    ]
    \Gamma_{k}
    \\
    & =
    \sum_{k}
    \Gamma_{k+q}^{\dag}
    [
        i E^{+} ( \alpha^{y}_{k,q} \eta_{x} -
        \alpha^{x}_{k,q} \eta_{y} )
    ]
    \Gamma_{k}
    + \cdots ,
\end{split}
\end{align}

where $E^{\pm}_{k,q} = E_{k} \pm E_{k+q}$. Note that the term
$E_{k,q}^{-}$ goes to zero at $q = 0$ and we have neglected it in the
final equality. The non-interacting terms in \eqref{eqn:EOM} can then be
grouped as

\begin{align}
    [ \xi_{q}, H_{0} ] - \omega \xi_{q} & =
    \sum_{k}
    \Gamma_{k+q}^{\dag}
    [ \bm{\alpha}'_{k,q} \cdot \bm{\eta} ]
    \Gamma_{k} ,
    \\
    \bm{\alpha}'_{k,q} & = R \bm{\alpha}_{k,q} =
    \begin{bmatrix}
        - \omega & i E^{+}_{k,q} & 0 \\
        -i E^{+}_{k,q} & - \omega & 0 \\
        0 & 0 & - \omega
    \end{bmatrix}
    \bm{\alpha}_{k,q} .
\end{align}

We now choose $\bm{\alpha}'$ such that $[\xi_{q}, H_{0}] - \omega \xi_{q}$ has the form

\begin{align}\label{eqn:EOMNonInteracting}
    [\xi_{q}, H_{0}] - \omega \xi_{q} =
    \begin{bmatrix}
        \gamma_{+} \\
        \gamma_{-}
    \end{bmatrix}
    \cdot
    \begin{bmatrix}
        \rho_{q}^{+} \\
        \rho_{q}^{-}
    \end{bmatrix} ,
\end{align}

for some constants $\gamma_{\pm}$ that we will solve for below. Using
the expression for $\rho_{q}^{\pm}$ (Eqs. \eqref{eqn:sigmaQ} and
\eqref{eqn:etaPMTransformed}) we find that $\bm{\alpha}'_{k,q}$ should be

\begin{align}\label{eqn:alphap}
    \bm{\alpha}'_{k,q} =
    \begin{bmatrix}
          \beta^{x}_{k,q} ( \gamma_{+} + \gamma_{-} ) \\
        i \beta^{y}_{k,q} ( \gamma_{+} - \gamma_{-} ) \\
          \beta^{z}_{k,q} ( \gamma_{+} + \gamma_{-} )
    \end{bmatrix} .
\end{align}

Thus, given the parameters $\gamma_{\pm}$ the form of $\xi_{q}$ is
fixed. We compute $\bm{\alpha}_{k,q}$ appearing in Eq. \eqref{eqn:xiqDef} by
inverting the matrix $R$:

\begin{align}
    \bm{\alpha}_{k,q} & = R^{-1} \bm{\alpha}'_{k,q} ,
    \\
    \label{eqn:RInverse}
    R^{-1} & =
    \begin{bmatrix}
        - \frac{\omega}{\omega^{2}  - (E_{k,q}^{+})^{2}} & 
        - \frac{i E_{k,q}^{+}}{\omega^{2} - (E_{k,q}^{+})^{2}}
        & 0
        \\
        \frac{i E_{k,q}^{+}}{\omega^{2}  - (E_{k,q}^{+})^{2}} &
        - \frac{\omega}{\omega^{2} - (E_{k,q}^{+})^{2}} &
        0
        \\
        0 &
        0 &
        - \frac{1}{\omega}
    \end{bmatrix} .
\end{align}

Next, we compute the commutator of $\xi_{q}$ with the
interaction Hamiltonian:

\begin{align}
\begin{split}
    [ \xi_{q}, H_{\intt} ] =
    \frac{U_{0}}{4}
    \sum_{Q}
    \left(
    [ \xi_{q}, \rho_{Q}^{+} ]
    \rho_{-Q}^{-}
    +
    \rho_{Q}^{+}
    [ \xi_{q}, \rho_{-Q}^{-} ]
    \right) .
\end{split}
\end{align}

We will evaluate this expression in the RPA limit. First, we set $Q = -
q$ in the first term and $Q = + q$ in the second (see Eq.
\eqref{eqn:baseCommutator}, this ensures that the $\Gamma_{k}$ operator
is only ever paired with $\Gamma_{k}^{\dag}$ at the same momentum).
Second, we replace $[\xi_{q}, \rho_{-q}^{\pm}]$ with its average,
$\langle \cdot \rangle$, over the BCS ground state. The result is

\begin{align}\label{eqn:hintRPA}
    [ \xi_{q}, H_{\intt} ] =
    \frac{U_{0}}{4}
    \left(
    \langle [ \xi_{q}, \rho_{-q}^{+} ] \rangle
    \rho_{q}^{-}
    +
    \langle [ \xi_{q}, \rho_{-q}^{-} ] \rangle
    \rho_{q}^{+}
    \right) .
\end{align}

Using Eq. \eqref{eqn:baseCommutator} we find

\begin{align}\label{eqn:hIntCommutator1}
\begin{split}
    [\xi_{q}, \rho_{-q}^{\pm}] =
    \sum_{k}
    \Gamma_{k}^{\dag}
    [ &
        (\bm{\alpha}_{k-q,q} \cdot \eta)
        (\bm{\beta}_{k,-q}^{\pm} \cdot \eta)
        -
    \\ &
        (\bm{\beta}_{k+q,-q}^{\pm} \cdot \eta)
        (\bm{\alpha}_{k,q} \cdot \eta)
    ]
    \Gamma_{k} ,
\end{split}
\end{align}

where $\bm{\beta}^{\pm}_{k,q} \cdot \bm{\eta} = M_{k+q}^{\dag}
\eta_{\pm} M_{k}$ was defined in Eq. \eqref{eqn:etaPMTransformed}.
Taking the ground state average is straightforward. The only operator
with a non-zero average is $(\Gamma_{k}^{\dag})_{2} (\Gamma_{k})_{2}$,
which comes from the lower right-hand corner of the matrix and has unit
average (independent of momentum $k$). Thus, to perform the ground
state average we simply compute the matrix in between the $\Gamma$
operators and take the lower-right matrix element. Doing this, and
shifting $k \rightarrow k + q$ in the integrand where appropriate we
write the above as a commutator between $2 \times 2$ matrices:

\begin{align}
\begin{split}
    \langle
        [\xi_{q}, \rho_{-q}^{\pm}]
    \rangle
    = \sum_{k}
    [ &
        \bm{\alpha}_{k,q} \cdot \eta,
        \bm{\beta}_{k+q,-q}^{\pm} \cdot \eta
    ]_{2,2}
    \\
    = \sum_{k}
    [ &
        \bm{\alpha}_{k,q} \cdot \eta,
        \bm{\beta}_{k,q}^{\pm} \cdot \eta
    ]_{2,2} .
\end{split}
\end{align}

The second equality follows from the fact that the $\beta^{0} \eta_{0}$
term naturally drops out and from a basic property of the $\alpha$ and
$\beta$ functions: $\beta_{k+q,-q}^{i} = \beta_{k,q}^{i}$ and
$\alpha_{k+q,-q}^{i} = \alpha_{k,q}^{i}$ for $i \in (x, \ y, \ z)$. We
can rewrite the relevant matrices as

\begin{align}
    \bm{\beta}_{k,q}^{\pm} \cdot \eta
    & =
    u_{k+q} u_{k} \eta_{\pm} - v_{k+q} v_{k} \eta_{\mp} + \beta_{k,q}^{z} \eta_{z} ,
    \\
    \bm{\alpha}_{k,q} \cdot \eta
    & =
    \alpha_{k,q}^{+} \eta_{+} +
    \alpha_{k,q}^{-} \eta_{-} + \alpha_{k,q}^{z} \eta_{z} .
\end{align}

Only $[\eta_{+}, \eta_{-}] = 4 \eta_{z}$ contributes to the $(2,2)$
component of the desired commutator.

\begin{align}
    \langle
        [\xi_{q}, \rho_{-q}^{\pm}]
    \rangle
    = \pm 4 \sum_{k} \bigl(
       \alpha_{k,q}^{\pm} v_{k+q} v_{k} +
       \alpha_{k,q}^{\mp} u_{k+q} u_{k} \bigr) .
\end{align}

Using the expression for $\bm{\alpha}'$ (Eq. \eqref{eqn:alphap}) and
using $\bm{\alpha} = R^{-1} \bm{\alpha}'$ (Eq. \eqref{eqn:RInverse}) we
find expressions for $\alpha^{\pm}$:

\begin{align}
    \alpha_{k,q}^{\pm} & =
    \frac{1}{\omega \pm E_{k,q}^{+}}
    (
        - \gamma_{\pm} u_{k+q} u_{k} + \gamma_{\mp} v_{k+q} v_{k}
    ) .
\end{align}

Putting these expressions into the commutators we find

\begin{align}
    \langle
        [\xi_{q}, \rho_{-q}^{+}]
    \rangle
    & =
    4 B \gamma_{+} +  4 A_{-} \gamma_{-} ,
    \\
    \langle
        [\xi_{q}, \rho_{-q}^{-}]
    \rangle
    & =
    4 A_{+} \gamma_{+} + 4 B \gamma_{-} ,
    \\
    B & = \sum_{k}
    2 E_{k,q}^{+} \frac{u_{k+q} u_{k} v_{k+q} v_{k}}{\omega^{2} - (E_{k,q}^{+})^{2}} ,
    \\
    A_{\pm} & = \pm \sum_{k}
    \frac{u_{k+q}^{2} u_{k}^{2} ( \omega \mp E_{k,q}^{+} )}
    {\omega^{2} - (E_{k,q}^{+})^{2}}
    \nonumber
    \\ & \quad \mp \sum_{k}
    \frac{v_{k+q}^{2} v_{k}^{2} ( \omega \pm E_{k,q}^{+} )}
    {\omega^{2} - (E_{k,q}^{+})^{2}} .
\end{align}

The term in $A_{\pm}$ which is linear in $\omega$ integrates to zero
and so

\begin{align}
    A_{\pm} = A =
    \sum_{k}
    - E_{k,q}^{+}
    \frac{u_{k+q}^{2} u_{k}^{2} + v_{k+q}^{2} v_{k}^{2}}
    {\omega^{2} - (E_{k,q}^{+})^{2}} .
\end{align}

Finally, we can put our expression for $\langle [\xi_{q},
\rho_{-q}^{\pm}] \rangle$ into Eq. \eqref{eqn:hintRPA} to obtain $[
    \xi_{q}, H_{\intt} ]$:

\begin{align}
    [ \xi_{q}, H_{\intt} ] & =
    U_{0}
    [
    ( \gamma_{+} A + \gamma_{-} B ) \rho_{q}^{+} +
    ( \gamma_{+} B + \gamma_{-} A ) \rho_{q}^{-}
    ]
    \\ & =
    U_{0}
    \left(
        \begin{bmatrix}
            A & B \\
            B & A
        \end{bmatrix}
        \begin{bmatrix}
            \gamma_{+} \\ \gamma_{-}
        \end{bmatrix}
    \right)
    \cdot
    \begin{bmatrix}
        \rho_{q}^{+} \\ \rho_{q}^{-}
    \end{bmatrix} .
\end{align}

Combining this with our expression for $[ \xi_{q}, H_{0} ] - \omega
\xi_{q}$ (Eq. \eqref{eqn:EOMNonInteracting}) gives

\begin{align}
    [ \xi_{q}, H ] - \omega \xi_{q} & =
    \left(
    \begin{bmatrix}
        \gamma_{+} \\ \gamma_{-}
    \end{bmatrix}
    +
    U_{0}
        \begin{bmatrix}
            A & B \\
            B & A
        \end{bmatrix}
        \begin{bmatrix}
            \gamma_{+} \\ \gamma_{-}
        \end{bmatrix}
    \right)
    \cdot
    \begin{bmatrix}
        \rho_{q}^{+} \\ \rho_{q}^{-}
    \end{bmatrix}
    + \cdots .
\end{align}

Now if the vector $(\gamma_{+}, \gamma_{-})$ is an eigenvector of the
$A$-$B$ matrix then the right-hand side is zero and the equations
of motion are satisfied when

\begin{align}
    \begin{bmatrix}
        A & B \\
        B & A
    \end{bmatrix}
    \begin{bmatrix}
        \gamma_{+} \\ \gamma_{-}
    \end{bmatrix}
    & =
    \lambda
    \begin{bmatrix}
        \gamma_{+} \\ \gamma_{-}
    \end{bmatrix} ,
    \\
    1 + U_{0} \lambda = 0 .
\end{align}

The eigenvalues and eigenvectors are

\begin{align}
    \lambda & = A + B,
    &&
    \begin{bmatrix}
        \gamma_{+} \\ \gamma_{-}
    \end{bmatrix}
    =
    \begin{bmatrix}
        1 \\ 1
    \end{bmatrix} ,
    \\
    \lambda & = A - B,
    &&
    \begin{bmatrix}
        \gamma_{+} \\ \gamma_{-}
    \end{bmatrix}
    =
    \begin{bmatrix}
        + 1 \\ -1
    \end{bmatrix} .
\end{align}

The two solutions here correspond to two collective modes. The values
of $\gamma_{\pm}$ found here completely determine the form of the
$\xi_{q}$ operator. Each eigenvalue gives a different condition on
$\omega$, it being the only remaining free parameter: $1 + U_{0} ( A
\pm B) = 0$. This expands to

\begin{align}
\begin{split}
    & 1 + U_{0} \sum_{k}
    E_{k,q}^{+}
    \frac{-(u_{k+q}^{2} u_{k}^{2} + v_{k+q}^{2} v_{k}^{2})
    \pm 2 u_{k+q} u_{k} v_{k+q} v_{k} }
    {\omega^{2} - (E_{k,q}^{+})^{2}}
    \\ & \qquad \qquad \qquad \qquad \qquad \qquad = 0 ,
\end{split}
\end{align}

which simplifies to

\begin{align}
    1 + U_{0} \sum_{k}
    \frac{E + E'}{2 E E'}
    \frac{ - E E' - \varepsilon \varepsilon' \pm \Delta^{2} }{\omega^{2} - (E + E')^{2}}
    = 0 .
\end{align}

Here, every primed quantity in the integrand is evaluated at $k+q$ while every
un-primed quantity is evaluated at $k$. This implicit equation for $\omega =
\omega(q)$ is the condition for a pole in the dynamic pair susceptibility (as in
Ref. \cite{littlewood_amplitude_1982}, and as was re-derived in Eq.
\eqref{eqn:poleCondition} of the main text). Note that the $+$ sign corresponds
to $\lambda = A + B$ and gives the dispersion relation for the Higgs mode, while
$-$ corresponds to $\lambda = A - B$ and gives the Goldstone dispersion. Note
that we have solved the equation of motion under two approximations: (i) we use
the RPA, and (ii) we neglect small terms containing $E^{-}_{k,q}$ and
$\beta^{0}_{k,q}$.

Finally, we wish to point out that our solution for $(\gamma_{+}, \gamma_{-})$
means that $[\xi_{q}, H_{0}] - \omega \xi_{q}$ is equal to $2 \rho_{q}^{x}$ for
the Higgs mode and $2 i \rho_{q}^{y}$ for the Goldstone mode.

\subsection{Explicit annihilation operator for the Higgs mode}

In addition to the dispersion relations, we also obtain the form of the
operator $\xi_{q}$ in terms of Fermionic operators. For the Higgs mode
we set $\gamma_{+} = \gamma_{-} = 1$. The resulting annihilation
operator is

\begin{align}
    \xi_{q} = & \sum_{k}
    \Psi_{k+q}^{\dag} \Xi_{k,q} \Psi_{k} ,
    \\
\begin{split}
\label{eqn:HiggsCreationKernel}
    \Xi_{k,q} = &
    M_{k+q}
    ( \bm{\alpha} \cdot \eta )
    M_{k}^{\dag}
    \\
    = &
    2 \frac{u'u-v'v}{\omega_{q} + E + E'}
    \begin{bmatrix}
        u'v   & -u'u \\
        v'v   & -v'u
    \end{bmatrix}
    \\ + &
    2 \frac{u'u-v'v}{\omega_{q} - E - E'}
    \begin{bmatrix}
        v'u  & v'v \\
        -u'u & -u'v
    \end{bmatrix}
    +
    \delta \Xi_{k,q} .
\end{split}
\end{align}

We have separated out a part $\delta \Xi_{k,q}$ given by

\begin{align}
\begin{split}
    \delta \Xi_{k,q} = & - 2
    \frac{u'v + v'u}{\omega_{q}}
    M_{k+q}
    \eta_{z}
    M_{k}^{\dag}
    \\
    = &
    -
    \frac{2}{\omega_{q}}
    (u'v + v'u)^{2} \eta_{x}
    -
    \frac{2}{\omega_{q}}
    \frac{\Delta(\varepsilon + \varepsilon')}{2 E E'}
    \eta_{z} .
\end{split}
\end{align}

At $q = 0$ the matrix in Eqn. \ref{eqn:HiggsCreationKernel} reduces to
the result obtained in Ref. \cite{tsuchiya_hidden_2018} obtained using
a Holstein-Primakoff transformation in the Anderson pseudo-spin picture
\emph{up to the term $\delta \Xi_{k,q}$} (the additional term in our
calculation is $\delta \Xi_{k,0} = - H_0(k) / E^{2}_{k}$). The
technique used in that work seems to be applicable only at $q = 0$,
where as here we produce the full $q$-dependence of the creation
operator. Note that the ``additional'' part $\delta \Xi_{k,q}$ is not
present for the Goldstone mode: in the Goldstone case the creation
operator we obtain reduces exactly to that in
\cite{tsuchiya_hidden_2018} in the $q \rightarrow 0$ limit.

\subsection{Interaction terms}

In solving the equations of motion we neglected small terms that were
proportional to $E^{-}_{k,q}$ and $\beta^{0}_{k,q}$. We now wish to
include these terms (that is, the $\cdots$ in Eq. \eqref{eqn:EOM}). The
only term of this kind which survives for the Higgs mode is

\begin{align}
    - \sum_{k}
    \Gamma_{k+q}^{\dag}
    E^{-}_{k,q} \alpha_{k,q}^{z} \eta_{0}
    \Gamma_{k} .
\end{align}

This can be reduced to

\begin{align}
    \frac{2}{\omega_{H,q}} \sum_{k}
    E^{-}_{k,q} \beta_{k,q}^{z}
    \Gamma_{k+q}^{\dag}
    \eta_{0}
    \Gamma_{k} .
\end{align}

From this we can derive an interaction Hamiltonian. If we treat $[\xi_{q}, H]$
as functional differentiation of the Higgs Hamiltonian with respect to
$\xi_{q}^{\dag}$, then we can functionally integrate the above term to obtain
the term which belongs to the Hamiltonian itself (as, for example, $\omega_{q}
\xi_{q}$ in Eq. \eqref{eqn:EOM} integrates to $\sum_{q} \omega_{q}
\xi_{q}^{\dag} \xi_{q}$). The result is an interaction term given by

\begin{align}
    \sum_{q}
    \frac{2}{\omega_{H,q}}
    \xi_{q}^{\dag}
    \sum_{k}
    E^{-}_{k,q} \beta_{k,q}^{z}
    \Gamma_{k+q}^{\dag}
    \eta_{0}
    \Gamma_{k} .
\end{align}

Note that the pre-factor $2/\omega_{H,q}$ which depends on $q$ only is slightly
arbitrary since we have not normalized the $\xi_{q}$ operator. The Hermitian
conjugate of this term will also appear in the Hamiltonian; it follows from a
calculation of $[\xi_{q}^{\dag}, H]$. Importantly, this 3-point interaction has
exactly the same structure as the 3-point interaction derived in the main text
(Eqn. \ref{eqn:fullHiggsHam}) using a different method.

\section{Material Parameters}

The table below collects the material parameters used in Fig.
\ref{fig:condensationCriteria}. It is important to note that these parameters
represent a mapping onto a simple single-band BCS model. The numbers for NbN,
which vary substantially in the literature, are taken from a particular
experiment on thin films and are meant to be illustrative; we use values for S2
in Ref. \cite{hazra_superconducting_2016} with $\Lambda$ alone supplied by Ref.
\cite{chockalingam_superconducting_2008}. Figure \ref{fig:condensationCriteria}
is plotted with the experimentally controllable parameters (not including
$\bar{c}_{1}$) on the axes and with material parameters fixed. Thus, we do not
account for changes to the material parameters (for example $\lambda_{L}$) with
the sample geometry in Fig. \ref{fig:condensationCriteria}.

\begin{table*}
\caption{\label{tab:materials} Material parameters used in Fig.~\ref{fig:condensationCriteria}.}
\begin{ruledtabular}
\begin{tabular}{lcccc}

Material & $k_F$ ($10^{8}\,\mathrm{cm}^{-1}$) & $\varepsilon_F/\Delta$ &
$\Lambda/\Delta$ & $\lambda_{L}$ (nm) \\
\colrule

Pb &
1.58
\cite{ashcroft_solid_1976} &
$6.94\times10^{3}$
\cite{ashcroft_solid_1976,parks_superconductivity_2018,parks_superconductivity_2018-1} &
6.44
\cite{parks_superconductivity_2018,parks_superconductivity_2018-1} &
39
\cite{parks_superconductivity_2018} \\

Nb &
1.18
\cite{ashcroft_solid_1976} &
$3.47\times10^{3}$
\cite{ashcroft_solid_1976,parks_superconductivity_2018,parks_superconductivity_2018-1} &
15.58
\cite{parks_superconductivity_2018,parks_superconductivity_2018-1} &
44
\cite{parks_superconductivity_2018} \\

Al &
1.75
\cite{ashcroft_solid_1976} &
$6.43\times10^{4}$
\cite{ashcroft_solid_1976,parks_superconductivity_2018,parks_superconductivity_2018-1} &
200.38
\cite{parks_superconductivity_2018,parks_superconductivity_2018-1} &
49
\cite{parks_superconductivity_2018} \\

NbN &
1.48
\cite{hazra_superconducting_2016} &
$2.99\times10^{3}$
\cite{hazra_superconducting_2016} &
5.35
\cite{hazra_superconducting_2016,chockalingam_superconducting_2008} &
192
\cite{hazra_superconducting_2016} \\

\end{tabular}
\end{ruledtabular}
\end{table*}

\bibliography{bib.bib}
\end{document}